\documentclass[11pt, a4paper, logo, copyright]{googledeepmind}

\pdftrailerid{redacted}

\makeatletter
\renewcommand\bibentry[1]{\nocite{#1}{\frenchspacing\@nameuse{BR@r@#1\@extra@b@citeb}}}
\makeatother

\usepackage{kantlipsum, lipsum}
\usepackage{dsfont}
\usepackage{gdm-colors}
\usepackage[utf8]{inputenc}   
\usepackage{newunicodechar}   
\usepackage{wasysym,marvosym}
\usepackage{ulem}
\usepackage{hyperref}       
\usepackage{algorithmicx}
\usepackage{algpseudocode}
\usepackage{multirow}
\usepackage{geometry} 
\usepackage[rightcaption]{sidecap} 

\usepackage{tikz}
\usetikzlibrary{shapes.geometric, arrows.meta, positioning}
\usepackage{amsmath}
\usepackage{algorithm}
\usepackage{algpseudocode}
\usepackage{listings}
\usepackage{enumitem} 
\usepackage{lipsum}
\usepackage[most]{tcolorbox}
\definecolor{thinkcolor}{RGB}{227,196,144}
\definecolor{observecolor}{RGB}{153,201,227}
\definecolor{explorecolor}{RGB}{178,217,200}

\newcounter{caseexample}[section]
\renewcommand{\thecaseexample}{\arabic{caseexample}}

\newcounter{promptexample}[section]
\renewcommand{\thepromptexample}{\arabic{promptexample}}

\usepackage{array, multirow, tabularx, booktabs, makecell}

\newcommand{\assignmentQuestionName}{Question} 

\usepackage{booktabs}
\usepackage{arydshln}
\usepackage{dashrule}
\usepackage{twemojis}

\usepackage[authoryear, sort&compress, round]{natbib}

\usepackage{bbding}
\usepackage[T1]{fontenc}    
\usepackage{url}            
\usepackage{booktabs}       
\usepackage{nicefrac}       
\usepackage{microtype}      
\usepackage{amsmath}
\usepackage{graphicx}
\usepackage{multicol}
\usepackage[nameinlink]{cleveref}
\usepackage{bbm}
\usepackage{multirow}
\usepackage{soul}
\usepackage{float}
\usepackage{wrapfig}
\usepackage{blindtext}
\usepackage{tablefootnote}
\usepackage{amsfonts}
\usepackage[flushleft]{threeparttable}
\usepackage{colortbl}
\usepackage{mathtools}
\usepackage{bm}
\usepackage{CJKutf8}
\usepackage{makecell}
\usepackage{caption}
\usepackage{capt-of}
\usepackage{array}
\usepackage{calc}      
\usepackage{caption}   
\usepackage{subcaption}  
\usepackage[bottom]{footmisc}
\usepackage{fontawesome}
\usepackage{tabularx}        
\usepackage{siunitx}

\newcolumntype{C}{>{\centering\arraybackslash}X}
\newcolumntype{L}{>{\raggedright\arraybackslash}X}

\title{RecGPT-V2 Technical Report}

\author{RecGPT Team}

\begin{abstract}
Large language models (LLMs) have demonstrated remarkable potential in transforming recommender systems from implicit behavioral pattern matching to explicit intent reasoning. While RecGPT-V1 successfully pioneered this paradigm by integrating LLM-based reasoning into user interest mining and item tag prediction, it suffers from four fundamental limitations: \textbf{(1)} computational inefficiency and cognitive redundancy across multiple reasoning routes; \textbf{(2)} insufficient explanation diversity in fixed-template generation; \textbf{(3)} limited generalization under supervised learning paradigms; and \textbf{(4)} simplistic outcome-focused evaluation that fails to match human standards.

To address these challenges, we present RecGPT-V2 with four key innovations. First, a \textbf{Hierarchical Multi-Agent System} restructures intent reasoning through coordinated collaboration, eliminating cognitive duplication while enabling diverse intent coverage. Combined with \textit{Hybrid Representation Inference} that compresses user-behavior contexts, our framework reduces GPU consumption by 60\% and improves exclusive recall from 9.39\% to 10.99\%. Second, a \textbf{Meta-Prompting} framework dynamically generates contextually adaptive prompts, improving explanation diversity by +7.3\%. Third, \textbf{constrained reinforcement learning} mitigates multi-reward conflicts, achieving +24.1\% improvement in tag prediction and +13.0\% in explanation acceptance. Fourth, an \textbf{Agent-as-a-Judge} framework decomposes assessment into multi-step reasoning, improving human preference alignment. 
Online A/B tests on Taobao demonstrate significant improvements: +2.98\% CTR, +3.71\% IPV, +2.19\% TV, and +11.46\% NER. RecGPT-V2 establishes both the technical feasibility and commercial viability of deploying LLM-powered intent reasoning at scale, bridging the gap between cognitive exploration and industrial utility.
\end{abstract}

\begin{document}
\begin{CJK*}{UTF8}{gkai}

\maketitle

\begin{figure}[h]
    \centering
    \includegraphics[width=\linewidth]{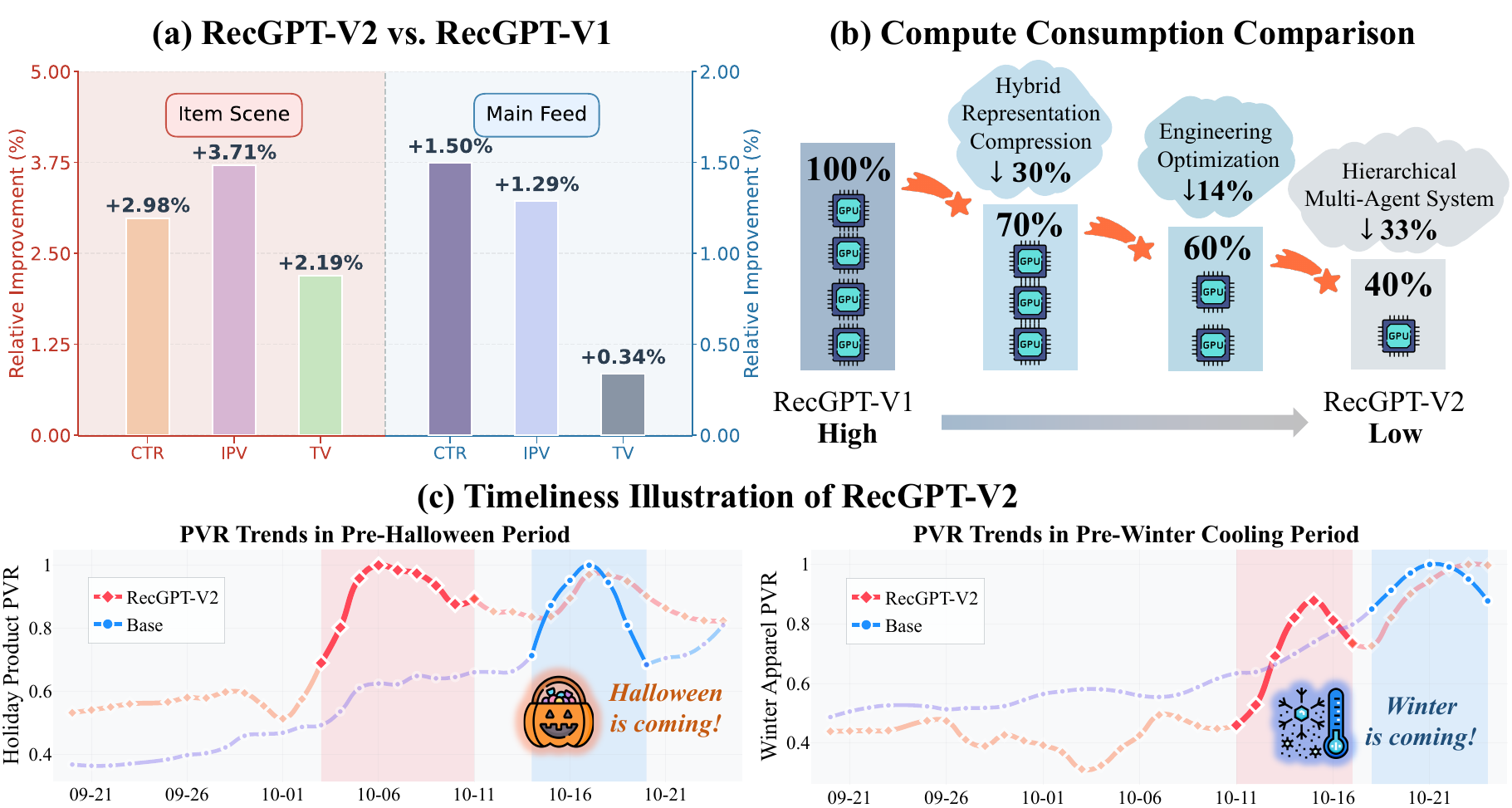}
    \caption{Comparison between RecGPT-V2 and RecGPT-V1: \textbf{(a)} online A/B performance in Taobao’s ``Guess What You Like''; \textbf{(b)} 60\% compute savings via optimization pipeline; and \textbf{(c)} Superior timeliness in capturing seasonal trends, exemplified by PVR trends for Halloween and Winter-related products.}
\end{figure}

\newpage
\setcounter{tocdepth}{2} 

\tableofcontents 

\newpage

\section{Introduction}
\label{sec:introduction}
Recommender systems have evolved significantly over the past two decades, progressing from matrix factorization~\citep{koren2009matrix} to deep neural networks~\citep{tang2025think}. Despite these advances, contemporary industrial systems remain fundamentally constrained by their reliance on historical behavioral patterns and log-fitting objectives, optimizing for behavioral pattern matching without explicitly reasoning about underlying user intent.
To address these issues, RecGPT-V1~\citep{yi2025recgpt} emerged as a paradigm-shifting framework that elevates user intent from implicit behavioral signals to explicit reasoning objectives. By integrating large language models (LLMs) into key stages, such as user interest mining and item tag prediction, RecGPT-V1 transforms traditional pattern matching into an intent-centric recommendation objective grounded in semantic understanding and logical reasoning.
This analytical reasoning paradigm enables the decomposition of complex recommendation tasks into interpretable and modular stages, facilitating transparent and controllable mapping from user intent understanding to item relevance prediction.

While RecGPT-V1 successfully leverages LLM-based knowledge and reasoning to improve recommendation quality and demonstrates promising online performance in industrial deployment, it still exhibits several limitations that hinder its scalability, efficiency, and effectiveness:

\textbf{Limitation 1: Computational inefficiency and redundant intent reasoning in multi-route architectures.}
RecGPT-V1 adopts a multi-route LLM-based channel\footnote{Following the initial RecGPT-V1 deployment, we extended the cognitive channel into multiple LLM-based retrieval routes, each specialized in leveraging different contextual and side information (\textit{e.g.}, weather, trending events, seasonal factors) to enhance semantic coverage and situational awareness in intent reasoning.} in which multiple LLM-based reasoning routes independently analyze user intent and retrieve item candidates.
Although this architecture broadens the semantic and contextual scope of user modeling, it exhibits substantial redundancy in both representation encoding and cognitive processes. \textbf{At the representation level}, each route encodes the full user behaviors sequence (averaging \textcolor{teal}{\textbf{32K}} tokens) even though only a small subset is relevant to the current intent prediction, leading to excessive computational overhead from repeatedly processing long sequences. \textbf{At the cognitive level}, different routes may demonstrate redundant reasoning outputs, generating overlapping recommendation candidates with an inter-route duplication rate reaching \textbf{\textcolor{teal}{13.46\%}}. Collectively, these inefficiencies result in significant computational waste and limit the overall scalability of RecGPT-V1 in large-scale industrial recommender systems.

\textbf{Limitation 2: Insufficient explanation diversity in the manner of fixed prompt templates.}
RecGPT-V1 employs fixed prompt templates to generate recommendation explanations by combining user interests and item attributes. This static approach produces homogeneous explanations that fail to capture the multi-dimensional and dynamic nature of personalized user needs. The templates cannot adaptively incorporate real-time contextual signals, resulting in generic explanations with limited personalization that struggle to engage users across diverse scenarios.

\textbf{Limitation 3: Supervised learning on static data limits generalization in complex generation tasks.}
RecGPT-V1 relies on supervised fine-tuning over curated high-quality corpora to learn recommendation-oriented generation tasks. This paradigm facilitates efficient transfer of human expertise but also anchors the model to fixed data distributions and explicit objective signals. In real-world recommendation scenarios, user needs evolve dynamically and involve multiple, interacting objectives with diverse operational constraints such as diversity, novelty, and relevance. Learning on static corpora cannot adequately capture these dynamic dependencies, resulting in limited generalization and unstable performance in multi-objective and multi-constraint generation tasks.

\textbf{Limitation 4: Simplistic outcome-focused evaluation in LLM-as-a-Judge.}
RecGPT-V1 employs LLM-as-a-Judge for one-shot outcome evaluation, training the judge to directly predict quality scores from data-label pairs. This result-oriented training paradigm collapses the multi-dimensional and multi-step reasoning inherent in human evaluation into a single prediction objective. By overlooking the intermediate reasoning steps that human evaluators employ to assess quality across multiple criteria (\textit{e.g.}, relevance, diversity, coherence), this approach limits the judge's ability to capture nuanced quality distinctions and reduces alignment with human evaluation standards.

\begin{figure}
    \centering
    \captionsetup{justification=centering}
    \includegraphics[width=0.9\textwidth]{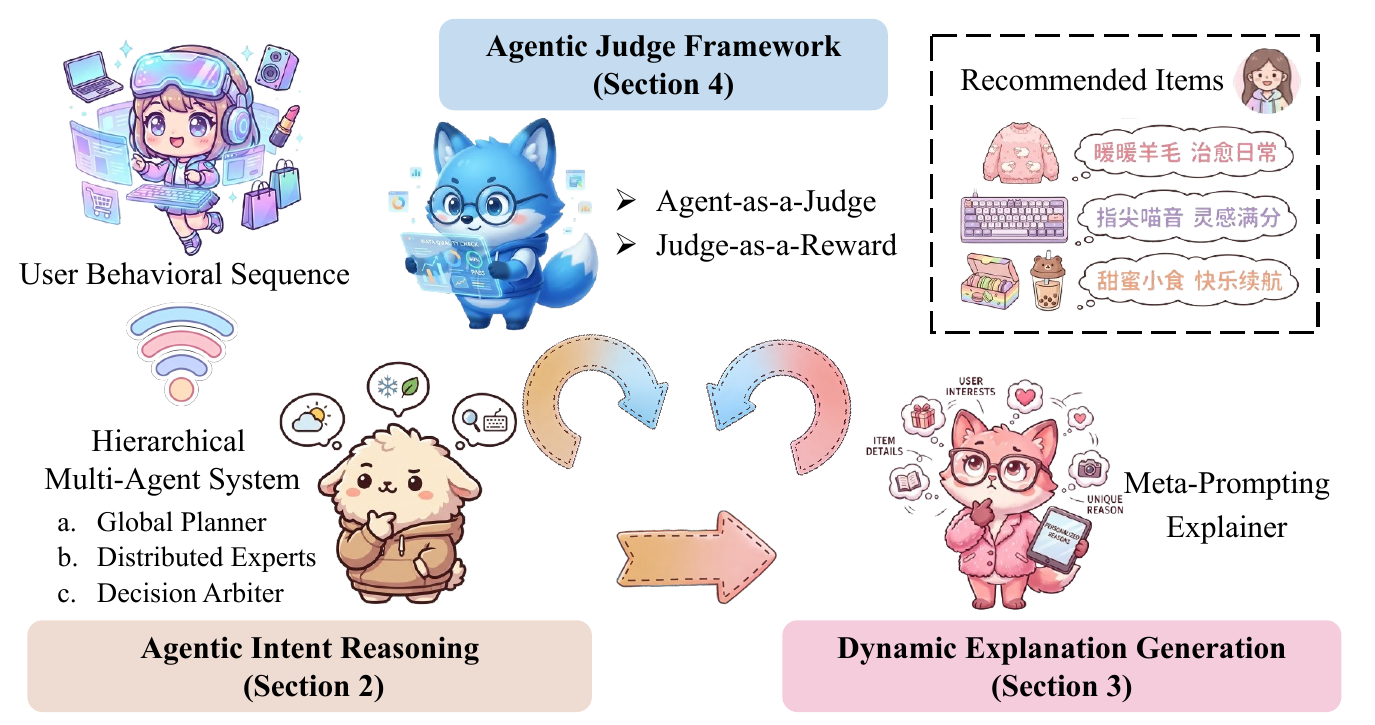}
    \caption{Overview of RecGPT-V2 architecture.}
    \label{fig:intro}
\end{figure}

To address these fundamental challenges, we present RecGPT-V2, which introduces three key architectural and algorithmic innovations:

\textbf{$\bigstar$ Agentic Intent Reasoning.}
To mitigate computational inefficiency and cognitive redundancy in multi-route architectures, we propose a \textbf{Hierarchical Multi-Agent System (HMAS)} that restructures LLM-based intent reasoning through coordinated multi-agent collaboration. By integrating multi-source environmental signals (\textit{e.g.}, trending events, weather patterns) into specialized expert agents, HMAS enables complementary reasoning across diverse contextual dimensions while eliminating cognitive duplication, increasing the exclusive recall from \textbf{\textcolor{teal}{9.39\%}} to \textbf{\textcolor{red}{10.99\%}}. To support efficient industrial deployment, we introduce \textit{Hybrid Representation Inference} that compresses user-behavior tokens from \textbf{\textcolor{teal}{32K}} to \textbf{\textcolor{red}{11K}} through atomized entity encoding, along with complementary \textit{Infrastructure Engineering Optimizations}. Together, these techniques improve MFU by \textbf{\textcolor{red}{+53.7\%}} and reduce GPU consumption by \textbf{\textcolor{red}{60.0\%}}, enabling scalable deployment without sacrificing reasoning quality.

\textbf{$\bigstar$ Dynamic Explanation Generation.}
To overcome the limitations of fixed prompt templates, we utilize a \textbf{Meta-Prompting} technique for dynamic recommendation explanation generation. By synthesizing user interests, item attributes, and real-time contextual signals (\textit{e.g.}, weather, seasonal events, trending topics), the meta-prompt generator autonomously constructs task-specific instruction templates that adapt to diverse scenarios and content characteristics. Compared to RecGPT-V1's static templates, our framework improves explanation diversity by \textbf{\textcolor{red}{+7.3\%}}, demonstrating that adaptive prompt engineering effectively enhances user engagement and satisfaction

\textbf{$\bigstar$ Constrained Reinforcement Optimization for Multi-Objective Generation.}
To overcome the limited generalization of supervised learning on static data, we propose a reinforcement-learning-based optimization framework for multi-objective recommendation generation tasks. Instead of directly summing multiple reward signals, we design a constrained reward shaping mechanism that guides the model to perform continual self-evolving within the feasible optimization domain.
Experiments show that our method improves human-evaluated tag quality pass rate by \textbf{\textcolor{red}{+24.0\%}} on item-tag prediction task, and increases the human-rated explanation acceptance rate by \textbf{\textcolor{red}{+77.6\%}} on recommendation explanation task compared with RecGPT-V1.

\textbf{$\bigstar$ Process-Oriented Multi-Step Evaluation.}
To address the limitations of outcome-focused evaluation, we propose an \textbf{Agent-as-a-Judge} framework that decomposes abstract assessment into structured multi-step reasoning. By progressively refining judgments across multiple dimensions (\textit{e.g.}, relevance, diversity, coherence) through iterative deliberation, this process-oriented paradigm enhances evaluation fidelity and aligns more closely with human standards. Experiments show that Agent-as-a-Judge outperforms LLM-as-a-Judge baselines, improving human preference alignment by \textbf{\textcolor{red}{+0.46\%}} on item tag prediction and \textbf{\textcolor{red}{+1.76\%}} on recommendation explanation generation, achieving near-human evaluation accuracy while retaining the cost-effectiveness of automated judging.

Figure~\ref{fig:intro} illustrates the overall architecture of RecGPT-V2. The system operates through a streamlined pipeline: lifelong user behaviors are compressed into hybrid contextual representations (\S\ref{sec:aec}), which feed into a Hierarchical Multi-Agent System for intent decomposition and item tag prediction (\S\ref{sec:hmas}). The predicted tags are grounded into in-corpus items through downstream recommenders, augmented with personalized explanations (\S\ref{sec:explanation}). To ensure generation quality and enable continuous improvement, we introduce an Agent-as-a-Judge evaluation framework (\S\ref{sec:agentic_judge}) for assessing generation tasks, coupled with a Judge-as-a-Reward distillation method (\S\ref{sec:judge_as_reward}) that transfers agent judgments into optimization reward signals.

In large-scale online A/B tests conducted on Taobao's homepage, RecGPT-V2 delivers significant performance improvements over RecGPT-V1, achieving a \textbf{\textcolor{red}{+3.64\%}} increase in IPV (Item Page Views), a \textbf{\textcolor{red}{+3.01\%}} lift in CTR (Click-Through Rate), a \textbf{\textcolor{red}{+2.11\%}} gain in TV (Transaction Volume), and a \textbf{\textcolor{red}{+11.46\%}} boost in NER (Novelty Exposure Rate).

\tcbset{
    promptbox/.code args={#1/#2}{
        \tcbset{
            enhanced,
            arc=0mm,
            colframe=#1, 
            colback=#2, 
            coltitle=black,
            fonttitle=\large\bfseries,
            attach boxed title to top left={xshift=0mm, yshift=-1.0mm},
            boxed title style={
                skin=enhancedfirst jigsaw,
                size=small,
                arc=3mm,
                bottom=0mm,
                left=8mm,
                right=8mm,
                top=1mm,
                colback=#1
            },
            boxrule=0pt,
            frame hidden,
            borderline north={4pt}{0pt}{#1},
        }
    }
}
\definecolor{exp_colback}{rgb}{0.949, 0.965, 0.980}
\definecolor{exp_colframe}{rgb}{0.878, 0.922, 0.965}

\section{Agentic Intent Reasoning}
\label{sec:multiagent}
As articulated in the introduction, RecGPT-V1's parallel multi-route cognitive architecture suffers from \textbf{dual-level computational inefficiency}: (1)~\textbf{representation-level waste}, where each route redundantly encodes the entire user behavior sequence (averaging \textbf{\textcolor{teal}{32K}} tokens) despite only a small fraction being relevant to its specific reasoning objective, and (2)~\textbf{cognitive-level overlap}, where isolated reasoning processes generate duplicated candidates, manifesting in a \textbf{\textcolor{teal}{13.46\%}} inter-route redundancy that squanders both processing resources and cognitive diversity.

To eliminate the above inefficiency issues, in this section, we propose a unified agentic framework that jointly improves \textbf{representation compactness} and \textbf{cognitive coordination}:
\begin{itemize}[topsep=0pt]
    \item \textbf{Hybrid Representation Inference} (\S\ref{sec:hri}): We propose a context compression method that achieves a \textbf{\textcolor{red}{7$\times$} compression ratio} by distilling behavior representations into single atomic units, dramatically reducing token length while preserving context integrity. 
    \item \textbf{Hierarchical Multi-Agent System} (\S\ref{sec:hmas}): We introduce a coordinated multi-agent architecture organized as Planner$\to$Experts$\to$Arbiter, which replaces isolated parallel routes with distributed collaborative reasoning. By integrating lifelong user behaviors and multi-source environmental signals (\textit{e.g.}, weather patterns and seasonal factors) into specialized expert agents, this design eliminates cognitive duplication while preserving diverse intent coverage.
\end{itemize}
Together, these innovations establish an efficient and scalable architecture for intent-driven recommendation at industrial scale.
In the following subsections, we elaborate on each component and provide respective analytical and empirical evaluations.

\subsection{Hybrid Representation Inference} \label{sec:hri}
Transformer-based LLMs exhibit computational complexity of $\mathcal{O}(L_{\text{in}}^2)$ in the prefill stage and $\mathcal{O}(L_{\text{in}} \times L_{\text{out}})$ in decoding, where $L_{\text{in}}$ and $L_{\text{out}}$ denote input/prompt and output/response lengths, respectively. In RecGPT-V1, user lifelong behaviors account for approximately \textbf{95.89\%} of input tokens, creating severe computation and memory bottlenecks that hinder scalability.
To address this challenge, we introduce: \textbf{(1)} \textbf{\textit{Atomized Entity Compression}} (\S\ref{sec:aec}) that distills behavioral entities into compact atomic representations, and \textbf{(2)} \textbf{\textit{Infrastructure Engineering Optimization}} (\S\ref{sec:infra}) with prefill--decode separation and kernel operator upgrade to meet industrial latency requirements.

\begin{figure}[t]
    \centering
    \includegraphics[width=\textwidth]{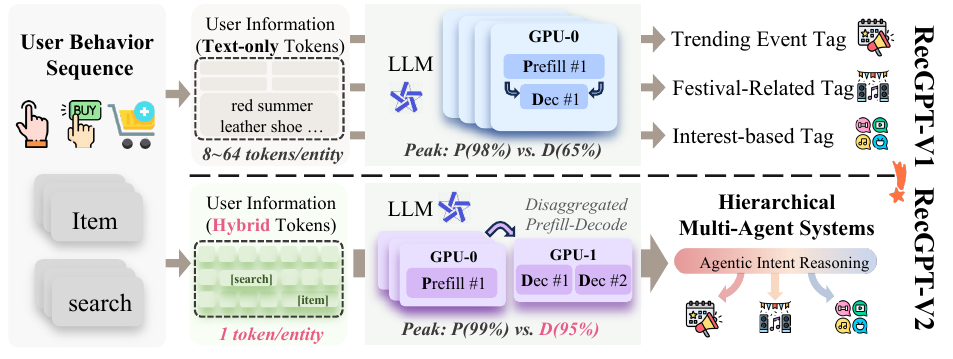}
    \caption{Comparison of inference architectures between RecGPT-V1 (full-text representation with coupled prefill-decode) and RecGPT-V2 (hybrid representation with disaggregated prefill-decode). RecGPT-V2 demonstrates substantial gains in GPU utilization peak and computational efficiency.}
    \label{fig:hri_framework_figure}
\end{figure}

\subsubsection{Atomized Entity Compression} \label{sec:aec}
The core principle underlying our approach is to compress entity information (including both item descriptions and user query histories) into atomic representational units, thereby substantially reducing context storage and computational overhead. This compression is achieved through a two-stage process: atomic representation encoding, and hybrid representation adaptation.

\paragraph{Stage 1: Atomic Representation Encoding} 
We employ pretrained embedding models (\textit{e.g.}, BGE~\citep{bge_embedding}, Qwen3-Embedding~\citep{zhang2025qwen3}, TBstars-Embedding) to encode entity information into dense vector representations. Formally, given an entity $e$ with its textual description $\mathbf{x} = [w_1, w_2, \ldots, w_n]$ consisting of $n$ tokens, we first obtain its embedding representation: 
\begin{equation*} 
\mathbf{h} = f_{\text{embed}}(\mathbf{x}) \in \mathbb{R}^{d_{\text{emb}}}, 
\end{equation*} 
where $f_{\text{embed}}(\cdot)$ denotes the embedding function that maps variable-length sequences to fixed-dimensional dense vectors, and $d_{\text{emb}}$ is the embedding dimension. To bridge the gap between the embedding space and the LLM's language space, we introduce a lightweight adaptor network $f_{\text{adapt}}(\cdot)$ that projects the embedding $\mathbf{h}$ into an atomic representation $\mathbf{z}$ compatible with LLM input:
\begin{equation*} 
\mathbf{z} = f_{\text{adapt}}(\mathbf{h}) = \mathbf{W}_2 \cdot \text{ReLU}(\mathbf{W}_1 \mathbf{h} + \mathbf{b}_1) + \mathbf{b}_2 \in \mathbb{R}^{d_{\text{LLM}}}, 
\end{equation*} 
where $\mathbf{W}_1 \in \mathbb{R}^{d_{\text{hidden}} \times d_{\text{emb}}}$, $\mathbf{W}_2 \in \mathbb{R}^{d_{\text{LLM}} \times d_{\text{hidden}}}$ are projection matrices, $\mathbf{b}_1, \mathbf{b}_2$ are bias terms, and $d_{\text{LLM}}$ matches the hidden dimension of LLMs. This atomic unit $\mathbf{z}$, denoted as \textcolor{red}{$[\text{entity}]$} in the context, replaces the original multi-token textual description. Case~\ref{case:entity_compression} illustrates a typical Chinese product title with 12 tokens compressed into a single atomic representation, achieving a 12:1 compression ratio.\footnote{As Taobao APP mainly targets Chinese users, all measures in this section (\textit{e.g.}, token counts, compression ratios) are computed on Chinese text. English translations in Case~\ref{case:entity_compression}, Case~\ref{case:sequence_compression} and Prompt~\ref{prompt:meta_prompt} are provided for better readability.}

\definecolor{case1_bg}{RGB}{245, 250, 255} 
\definecolor{case1_frame}{RGB}{52, 152, 219} 
\definecolor{token_bg}{RGB}{255, 248, 220} 
\definecolor{token_text}{RGB}{34, 139, 34} 
\definecolor{id_bg}{RGB}{240, 240, 240} 
\definecolor{id_text}{RGB}{25, 25, 112} 
\definecolor{entity_text}{RGB}{220, 20, 60} 

\refstepcounter{caseexample}\label{case:entity_compression}%
\noindent 
\begin{tcolorbox}[ 
enhanced, 
colback=case1_bg, 
colframe=case1_frame, 
boxrule=2pt, 
arc=4mm, 
width=\textwidth, 
title={\textbf{Case \thecaseexample: Entity Compression Example}}, 
coltitle=white, 
fonttitle=\bfseries, 
attach boxed title to top left={yshift=-3mm, xshift=6mm}, 
boxed title style={colback=case1_frame, arc=3mm}
] 
\vspace{0.1cm} 
\textcolor{case1_frame}{\textbf{Original Text:}}\;
情侣家居外穿防滑毛绒保暖可爱棉鞋\ /\ Couples' Indoor-Outdoor Wearable Non-Slip Plush Thermal Cotton Slippers

\vspace{0.1cm} 
\textcolor{token_text}{\textbf{Tokenized Sequence:}}\; 
\colorbox{token_bg}{\texttt{\textcolor{token_text}{情侣}}} 
\colorbox{token_bg}{\texttt{\textcolor{token_text}{家居}}} 
\colorbox{token_bg}{\texttt{\textcolor{token_text}{外}}} 
\colorbox{token_bg}{\texttt{\textcolor{token_text}{穿}}} 
\colorbox{token_bg}{\texttt{\textcolor{token_text}{防}}} 
\colorbox{token_bg}{\texttt{\textcolor{token_text}{滑}}} 
\colorbox{token_bg}{\texttt{\textcolor{token_text}{毛}}} 
\colorbox{token_bg}{\texttt{\textcolor{token_text}{绒}}} 
\colorbox{token_bg}{\texttt{\textcolor{token_text}{保暖}}} 
\colorbox{token_bg}{\texttt{\textcolor{token_text}{可爱}}} 
\colorbox{token_bg}{\texttt{\textcolor{token_text}{棉}}} 
\colorbox{token_bg}{\texttt{\textcolor{token_text}{鞋}}} 

\vspace{0.1cm} 
\textcolor{id_text}{\textbf{Token IDs:}}\; 
\colorbox{id_bg}{\texttt{\textcolor{id_text}{73245}}} 
\colorbox{id_bg}{\texttt{\textcolor{id_text}{49477}}} 
\colorbox{id_bg}{\texttt{\textcolor{id_text}{2382}}} 
\colorbox{id_bg}{\texttt{\textcolor{id_text}{8123}}} 
\colorbox{id_bg}{\texttt{\textcolor{id_text}{4153}}} 
\colorbox{id_bg}{\texttt{\textcolor{id_text}{11369}}} 
\colorbox{id_bg}{\texttt{\textcolor{id_text}{9144}}} 
\colorbox{id_bg}{\texttt{\textcolor{id_text}{44785}}} 
\colorbox{id_bg}{\texttt{\textcolor{id_text}{79318}}} 
\colorbox{id_bg}{\texttt{\textcolor{id_text}{28525}}} 
\colorbox{id_bg}{\texttt{\textcolor{id_text}{29441}}} 
\colorbox{id_bg}{\texttt{\textcolor{id_text}{18067}}} 

\vspace{0.1cm} 
\textcolor{entity_text}{\textbf{Compressed Atomic Representation:}}\; 
\texttt{[\textcolor{teal}{entity}]}\\
\colorbox{yellow!50}{\textbf{\textcolor{red}{(Compression Ratio is 12:1)}}}
\end{tcolorbox}

\textbf{Advantages of Adaptor-Based Projection.} Our approach offers three key advantages over existing methods (e.g., OneRec-Think~\citep{liu2025onerec}, LC-Rec~\citep{zheng2024adapting}, CoLLM~\citep{zhang2025collm}) that directly insert new tokens into the vocabulary of LLMs:
\begin{enumerate}[topsep=0pt, itemsep=2pt, label=\roman*)]
\item \textbf{Parameter Efficiency}: We only optimize the adaptor parameters while keeping the LLM backbone frozen, significantly reducing training cost and memory footprint.
\item \textbf{Superior Generalization}: By maintaining frozen LLM parameters, our approach preserves the model's original language understanding capabilities. The adaptor learns to project entities into the semantic space rather than forcing the model to recognize entirely new tokens. 
\item \textbf{Enhanced Modularity}: The decoupled design allows seamless integration with different embedding models and LLM architectures without modifying the base models. 
\end{enumerate}

\definecolor{case2_bg}{RGB}{252, 248, 245}  
\definecolor{case2_frame}{RGB}{191, 97, 106}  
\definecolor{original_box}{RGB}{255, 250, 240} 
\definecolor{compressed_box}{RGB}{240, 248, 255} 
\definecolor{action_color}{RGB}{163, 73, 164}  

This compression extends to complete user behavioral sequences. Case~\ref{case:sequence_compression} demonstrates a realistic scenario where a user profile with \textcolor{teal}{\textbf{21,349}} tokens is reduced to \textcolor{red}{\textbf{5,158}} tokens (token reduction ratio: \textbf{76\%}) by replacing item descriptions and query texts with atomic representations while preserving user attributes and temporal metadata with natural language. 
This hybrid representation effectively balances compactness and contextual richness.

\refstepcounter{caseexample}\label{case:sequence_compression}%
\noindent 
\begin{tcolorbox}[ 
enhanced, 
colback=case2_bg, 
colframe=case2_frame, 
boxrule=2pt, 
arc=4mm, 
width=\textwidth, 
title={\textbf{Case \thecaseexample: Complete User Behavioral Sequence Compression}}, 
coltitle=white, 
fonttitle=\bfseries, 
attach boxed title to top left={yshift=-3mm, xshift=6mm}, 
boxed title style={colback=case2_frame, arc=3mm}, 
breakable
] 
\vspace{0.1cm} 
\textcolor{blue!70!black}{\textbf{Original Full-Text Context}} (\textcolor{black}{\textbf{21,349 tokens}}): 

\vspace{0.2cm} 
\colorbox{green!5}{ 
\begin{minipage}{0.95\linewidth} 
\textbf{User Attributes:}\;
28岁女性,居住在北京市,双子座,属牛的用户\ /\ 28-year-old female resident of Beijing; Astrological signs: Gemini (Western), Ox (Chinese zodiac)

\vspace{0.2cm} 
\textbf{User Behavioral History:}\\ 
\textcolor{action_color}{3年前购买\ /\ Purchased 3 years ago} | \colorbox{yellow!30}{\texttt{高筒靴女秋冬新款}} \colorbox{yellow!30}{\texttt{缉线装饰丝缎质感连衣裙}}\ /\ \colorbox{yellow!30}{Women's autumn-winter knee-high boots} \colorbox{yellow!30}{Topstitched satin-textured dress}\\ 
\textcolor{action_color}{2年前搜索\ /\ Searched 2 years ago} | \colorbox{orange!30}{\texttt{高级感超好看外套}} \colorbox{orange!30}{\texttt{复古蓝牙小音箱}}\ /\ \colorbox{orange!30}{Premium aesthetic outerwear} \colorbox{orange!30}{Retro Bluetooth mini speaker} \\ 
\textcolor{action_color}{1年前点击\ /\ Clicked 1 year ago} | \colorbox{pink!30}{\texttt{韩版宽松毛衣}} \colorbox{pink!30}{\texttt{纯棉四件套}}\ /\ \colorbox{pink!30}{Korean-style loose-fit sweater} \colorbox{pink!30}{Pure cotton 4-piece bedding set}\\ 
\textcolor{gray}{\textit{$\vdots$ (numerous additional interactions omitted due to space)}} 
\end{minipage} 
} 
\begin{center} 
$\Downarrow$ \textbf{Atomized Entity Compression} 
\end{center} 
\textcolor{purple}{\textbf{Hybrid Representation Context}} (\textcolor{red}{\textbf{5,158 tokens}}): 

\vspace{0.2cm} 
\colorbox{compressed_box}{ 
\begin{minipage}{0.95\linewidth} 
\textbf{User Attributes:}\;
28岁女性,居住在北京市,双子座,属牛的用户\ /\ 28-year-old female resident of Beijing; Astrological signs: Gemini (Western), Ox (Chinese zodiac)

\vspace{0.2cm} 
\textbf{User Behavioral History:}\\ 
\textcolor{action_color}{3年前购买\ /\ Purchased 3 years ago} | \colorbox{yellow!30}{\texttt{[\textcolor{teal}{entity}]}} \colorbox{yellow!30}{\texttt{[\textcolor{teal}{entity}]}}\\ 
\textcolor{action_color}{2年前搜索\ /\ Searched 2 years ago} | \colorbox{orange!30}{\texttt{[\textcolor{teal}{entity}]}} \colorbox{orange!30}{\texttt{[\textcolor{teal}{entity}]}}\\ 
\textcolor{action_color}{1年前点击\ /\ Clicked 1 year ago} | \colorbox{pink!30}{\texttt{[\textcolor{teal}{entity}]}} \colorbox{pink!30}{\texttt{[\textcolor{teal}{entity}]}}\\ 
\textcolor{gray}{\textit{$\vdots$ (all other interactions similarly compressed)}} 
\end{minipage} 
} 

\vspace{0.1cm} 
\colorbox{yellow!50}{\textbf{\textcolor{red}{(Token Reduction: 76\%)}}} 
\end{tcolorbox}

However, the introduction of atomic units raises a critical question: 
\textit{How can we enable the LLM to seamlessly understand hybrid contexts that interleave natural language tokens with compressed entity representations?} 
To address this challenge, in the next section, we introduce a dedicated \textbf{Hybrid Representation Adaptation} to align the atomic units with the language space.

\paragraph{Stage 2: Hybrid Representation Adaptation}
To bridge this representational gap, we design a two-tier training strategy comprising \textbf{Self-Perception Tasks} and \textbf{Production-Oriented Alignment}. Importantly, during this adaptation phase, we \textcolor[RGB]{100,180,255}{keep the LLM backbone frozen} \twemoji{snowflake} and only \textcolor[RGB]{255,80,40}{train the adaptor parameters} \twemoji{fire}, ensuring parameter efficiency and preserving the model's pretrained general knowledge. Both training strategies share a unified formalization and optimization objective.

\textbf{(1) Self-Perception Tasks.} 
We adopt a ``\textit{what-is-it}'' philosophy to cultivate fine-grained entity understanding. Rather than relying on simple title reconstruction, we leverage powerful LLMs (\textit{e.g.}, GPT-4~\citep{achiam2023gpt}) to automatically generate diverse, attribute-focused questions that probe the semantic completeness of atomic representations. This dynamic question generation method follows a In-Context-Learning~\citep{brown2020language,dong2024survey} prompting strategy to ensure coverage of critical entity attributes. The meta-prompt design is illustrated in the Prompt~\ref{prompt:meta_prompt}.

\definecolor{meta_bg}{RGB}{252, 245, 255}
\definecolor{meta_frame}{RGB}{142, 68, 173}
\definecolor{example_in}{RGB}{225, 242, 252}
\definecolor{example_out}{RGB}{255, 243, 224}

\refstepcounter{promptexample}\label{prompt:meta_prompt}%
\noindent
\begin{tcolorbox}[
    enhanced,
    colback=meta_bg,
    colframe=meta_frame,
    boxrule=2pt,
    arc=4mm,
    width=\textwidth,
    title={\textbf{Prompt \thepromptexample: Meta-Prompt for Dynamic QA Pair Generation}},
    coltitle=white,
    breakable,
    fonttitle=\bfseries,
    attach boxed title to top left={yshift=-3mm, xshift=6mm},
    boxed title style={colback=meta_frame, arc=3mm}
]

\vspace{0.2cm}
\textbf{System Instruction:}\\
For a given product title, I want to verify whether the embedding model provides complete representational information. Please design corresponding questions and answers to confirm information completeness. All questions must be answerable from the input text alone. Output the result directly in JSON format without any additional text.

\vspace{0.3cm}
\colorbox{example_in}{\textbf{Example Input:}}\\[0.2cm]
{\small\texttt{情侣家居外穿防滑毛绒保暖可爱棉鞋}\\
Couples' Indoor-Outdoor Wearable Non-Slip Plush Thermal Cotton Slippers}

\vspace{0.2cm}
\colorbox{example_out}{\textbf{Example Output:}}
{\small\ttfamily
\begin{verbatim}
[
  {"Q": "What is the material of <entity>", "A": "Cotton"},
  {"Q": "What season is <entity> suitable for?", "A": "Winter"},
  {"Q": "What is the anti-slip performance of <entity>?", "A": "Non-slip"},
  {"Q": "What scenarios is <entity> suitable for?", "A": "Indoor&Outdoor"}
]
\end{verbatim}
}

\vspace{0.2cm}
\colorbox{orange!20}{\textbf{Actual Input:}}
{\small\texttt{澳洲进口美利奴羊毛半开拉链毛衣}}\\
Australian imported merino wool half-zip sweater

\vspace{0.2cm}
\colorbox{yellow!20}{\textbf{Generated Output:}}
\textit{(Model dynamically generates diverse attribute-focused QA pairs)}
\end{tcolorbox}

Formally, given an entity $e$ with original text $\mathbf{x}$, we leverage a powerful LLM (e.g., GPT-4) to automatically generate diverse attribute-focused question-answer pairs:
\begin{equation*}
\{(\mathbf{q}_i, \mathbf{a}_i)\}_{i=1}^K = \text{LLM}(\mathbf{x}),
\end{equation*}
where $K$ denotes the number of generated QA pairs. Each question $\mathbf{q}_i$ probes specific entity attributes, with the answer $\mathbf{a}_i$ extracted directly from $\mathbf{x}$. These QA pairs serve as supervision for training the adaptor to preserve semantic completeness in compressed representations.

\textbf{(2) Production-Oriented Alignment.}
To validate practical applicability and reinforce the adaptor's ability to project entity representations into semantically meaningful regions of the LLM's input space, we integrate compressed atomic units into two core recommendation generation tasks from RecGPT-V1, namely \textbf{User Interest Mining} and \textbf{Item Tag Prediction}:
\begin{itemize}[topsep=0pt]
    \item \textbf{User Interest Mining}: Infers user interest profiles from interaction histories, capturing both long-term preferences and short-term behavioral trends.
    \item \textbf{Item Tag Prediction}: Anticipates user intent by predicting relevant item tags based on inferred interests and historical behaviors.
\end{itemize}

\noindent
For each task, we first construct reference samples using full textual representations. Given a prompt containing complete entity descriptions, we obtain ground-truth responses from the frozen LLM, which serve as supervision signals for adaptor training.

\paragraph{Unified Training Formulation}
Both self-perception QA tasks and production-oriented tasks share an identical optimization paradigm. The core idea is to train the adaptor such that hybrid prompts (with compressed entities) can reproduce the same responses as full-text prompts would generate.

Formally, given any reference sample with full-text prompt $\mathcal{P}_{\text{full}}$ and its corresponding response $\mathbf{y}^*$, we construct its compressed counterpart by replacing all entity texts with adaptor-projected representations. The hybrid prompt is defined as:
\begin{equation}
\mathcal{P}_{\text{hybrid}} = \phi(\mathcal{P}_{\text{full}}), \quad \text{where } \phi(\mathbf{x}_e) = f_{\text{adapt}}(f_{\text{embed}}(\mathbf{x}_e)), \, \forall e \in \mathcal{E},
\end{equation}
where $\mathcal{E}$ denotes all entities in $\mathcal{P}_{\text{full}}$, and $\phi(\cdot)$ performs entity-to-atomic replacement. We optimize the adaptor to minimize the cross-entropy loss between model predictions on compressed inputs and reference responses, which is formulated as follows:
\begin{equation}
\mathcal{L}(\theta_{\text{adapt}}) = -\sum_{t=1}^{|\mathbf{y}^*|} \log p\left(y_t^* \mid \mathcal{P}_{\text{hybrid}}, \mathbf{y}_{<t}^*\right),
\end{equation}
where $p(\cdot)$ denotes the frozen LLM's output distribution and $\theta_{\text{adapt}}$ represents the adaptor parameters. This objective ensures that the adaptor learns semantic-preserving projections that maintain \textbf{functional equivalence} between compressed and full-text representations across diverse reasoning tasks.

The training corpus combines self-perception QA pairs and production task samples. Through joint optimization over these heterogeneous supervision signals, the adaptor achieves a \textbf{7$\times$} compression ratio while preserving task performance. Compared to vocabulary-expansion methods requiring full model fine-tuning, our strategy offers superior parameter efficiency and generalization capability.

\subsubsection{Infrastructure Engineering Optimization} \label{sec:infra}

To meet the stringent latency requirements of industrial-scale deployment, we introduce two complementary infrastructure optimizations that significantly enhance inference efficiency: \textbf{(1)} \textbf{\textit{Disaggregated Prefill-Decode Serving Architecture}} that strategically allocates computational resources according to phase-specific characteristics, and \textbf{(2)} \textbf{\textit{Advanced Kernel Integration with XQA Operators}} that leverage FP8 precision for accelerated attention computation on H20 GPUs.

\paragraph{Disaggregated Prefill-Decode Architecture}
Recommendation generation tasks exhibit a distinctive asymmetric input-output characteristic. Specifically, user behaviors and contextual information typically span \textbf{\textcolor{darkgray}{$\sim$10K tokens}}
, while outputs usually range from \textbf{\textcolor{darkgray}{hundreds of tokens}}. This results in an extreme input-to-output length ratio, creating substantial inefficiencies in traditional one-serving serving architectures where both prefill and decode phases execute on the same GPU resources, leading to suboptimal Model FLOPs Utilization (MFU) and limited throughput scalability.

The computational profiles of these two phases differ fundamentally:
\begin{itemize}[topsep=2pt, itemsep=2pt]
    \item \textbf{Prefill phase} is compute-intensive, processing extensive inputs through parallel attention mechanisms with complexity $\mathcal{O}(L_{\text{in}}^2)$. Moreover, once the KV cache is computed, it does not require persistent storage within the prefill worker and can be transferred to decode workers.
    \item \textbf{Decode phase} is memory-intensive, characterized by autoregressive generation with complexity $\mathcal{O}(L_{\text{in}} \times L_{\text{out}})$ and frequent KV cache accesses. The uncertain output lengths and sequential dependency make it inherently amenable to cache-based optimizations.
\end{itemize}

To improve resource utilization and computational efficiency, following prior work~\citep{zhong2024distserve,liu2024deepseek}, we adopt a disaggregated serving architecture that strategically partitions GPU resources according to phase-specific computational demands. We assign a \textbf{larger GPU pool to prefill operations} to maximize parallel throughput for long-context processing, while dedicating \textbf{fewer resources to decode operations} that primarily benefit from efficient memory access patterns. The two phases communicate through optimized KV cache transfer mechanisms, enabling each stage to operate at its optimal resource configuration.

\paragraph{XQA Kernel Integration}
To further optimize attention computation, we replace the previous FlashInfer kernel with XQA kernel to leverage FP8 precision inference on H20 GPUs. While FlashInfer is optimized primarily for BF16 precision, XQA kernel provides superior performance for \textbf{FP8 quantized models}, enabling faster attention computation with reduced memory bandwidth requirements.

\begin{wrapfigure}{r}{0.47\textwidth}
    \centering
    \vspace{-20pt}
    \includegraphics[width=0.9\linewidth]{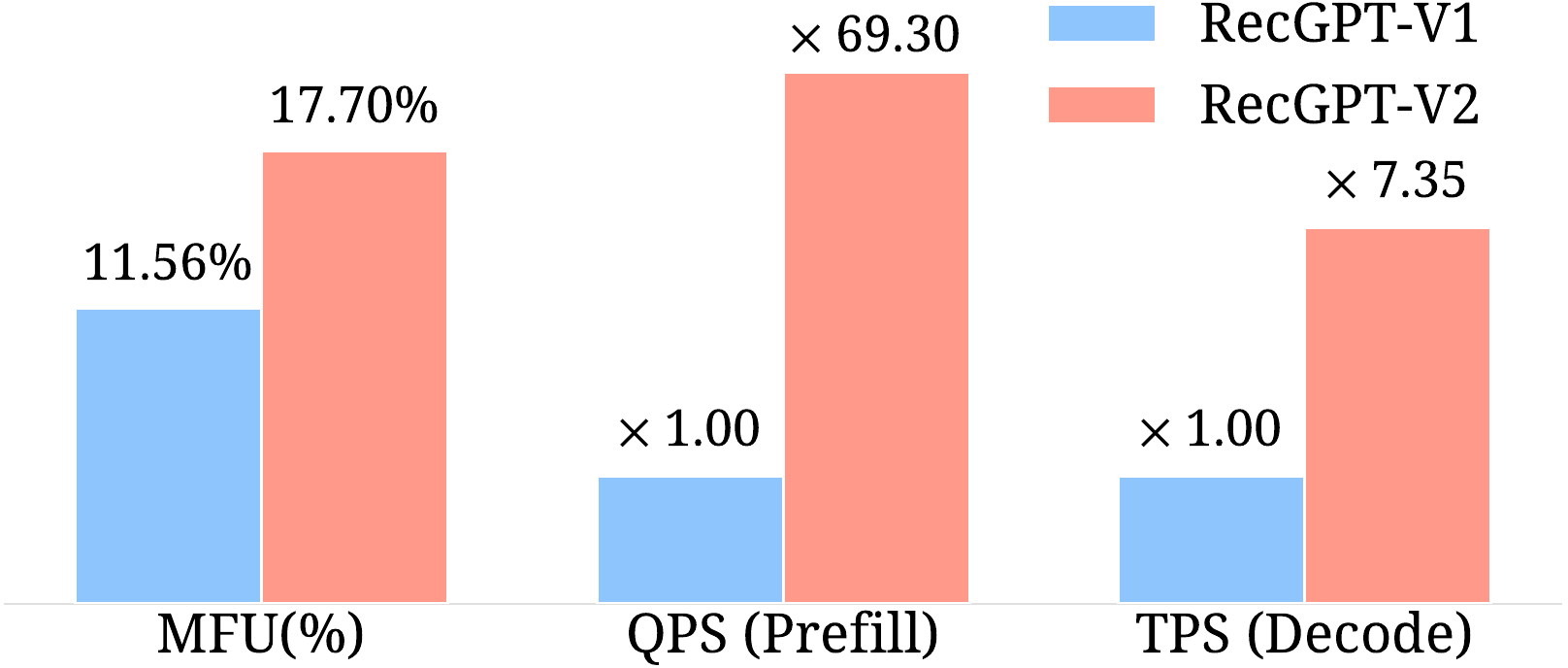}
    \vspace{-10pt}
    \caption{Computational efficiency comparison.}
    \label{fig:hmas_compute_efficiency}
    \vspace{-15pt}
\end{wrapfigure}

\paragraph{Performance Impact}
These infrastructure optimizations collectively improve overall MFU from \textbf{\textcolor{teal}{11.56\%}} (RecGPT-V1) to \textbf{\textcolor{red}{17.04\%}}. Combined with Atomized Entity Compression (\S\ref{sec:aec}) and the coordinated reasoning architecture from Hierarchical Multi-Agent System (\S\ref{sec:hmas}), RecGPT-V2 achieves a \textbf{\textcolor{red}{53.11\%}} improvement in MFU compared to RecGPT-V1. Furthermore, our system delivers substantial throughput gains with \textbf{\textcolor{red}{$\times$69.30}} QPS improvement in the prefill stage and \textbf{\textcolor{red}{$\times$7.35}} TPS improvement in the decode stage, enabling cost-effective scaling to industrial traffic volumes, as illustrated in Figure~\ref{fig:hmas_compute_efficiency}.

\subsection{Hierarchical Multi-Agent System} \label{sec:hmas}

Having established efficient representation inference through atomized entity compression and infrastructure optimizations, we now address the remaining inefficiencies in RecGPT-V1's isolated multi-route architecture. As mentioned in Section~\ref{sec:introduction}, the parallel reasoning routes independently encode identical user contexts and perform redundant cognitive processes, resulting in both \textbf{computational overhead from repeated full-sequence encoding} and \textbf{cognitive redundancy from overlapping predictions}, where the latter manifests in a \textbf{\textcolor{teal}{13.46\%}} inter-route duplication rate.

To jointly eliminate these dual-level inefficiencies, we propose a \textbf{Hierarchical Multi-Agent System (HMAS)} that restructures LLM-based intent reasoning into a coordinated three-tier architecture: \textbf{Planner $\to$ Experts $\to$ Arbiter}. The Global Planner decomposes user intent into specialized \textit{personas} by analyzing hybrid compressed context and multi-source contextual signals (\S\ref{sec:planner}). Each persona guides an expert agent to conduct role-specific item tag prediction, enabling parallel yet complementary reasoning without redundant full-context encoding (\S\ref{sec:experts}). The Decision Arbiter synthesizes expert predictions through collaborative reasoning (\S\ref{sec:arbiter}), producing refined candidate tags for downstream item retrieval. This design effectively eliminates both computational waste and cognitive duplication while preserving diverse intent coverage.

\begin{figure}[t]
    \centering
    \includegraphics[width=\textwidth]{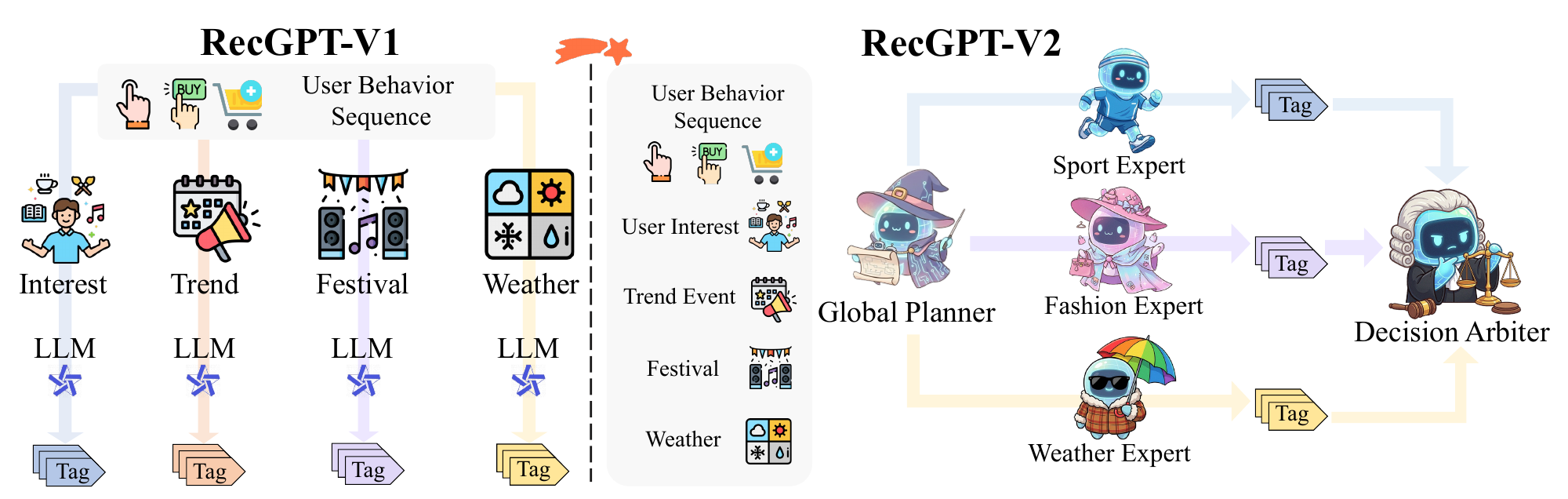}
    \vspace{-12pt}
    \caption{Architectural comparison between RecGPT-V1's isolated multi-route reasoning and RecGPT-V2's Hierarchical Multi-Agent System (Global Planner $\rightarrow$ Distributed Experts $\rightarrow$ Decision Arbiter), demonstrating reduced cognitive redundancy through coordinated intent decomposition.}
    \label{fig:hmas}

\end{figure}

\subsubsection{Global Planner}
\label{sec:planner}

The Global Planner serves as the top-level orchestrator in HMAS, responsible for decomposing complex user intent into a set of specialized \textit{personas} that guide downstream expert reasoning. Unlike RecGPT-V1's isolated parallel routes that independently process same contexts, the Global Planner performs holistic intent analysis by synthesizing rich contextual signals into a coherent strategic plan.

\paragraph{Context Representation}
The Global Planner receives a comprehensive contextual representation $\mathcal{C}$ comprising three complementary information sources:

\begin{enumerate}[topsep=0pt, label=(\roman*)]
\item \textbf{User Behavioral History} $\mathcal{B} = \{(a_i, e_i, t_i)\}_{i=1}^{N}$: Following RecGPT-V1, we aggregate chronologically ordered user interactions into temporally structured behavioral sequences, where each interaction is characterized by an action type $a_i \in \{\text{click}, \text{purchase}, \text{search}, \ldots\}$, an entity $e_i$ (item or query), and timestamp $t_i$. This temporal aggregation provides a compact yet informative representation of user engagement patterns. 

\item \textbf{User Profile} $\mathcal{U} = \{\mathcal{U}_{\text{attr}}, \mathcal{U}_{\text{int}}\}$: The user profile consists of two components:
\begin{itemize}[topsep=2pt, itemsep=2pt]
    \item Static Attributes $\mathcal{U}_{\text{attr}}$: Demographic information including age, gender, location, and other stable characteristics. 
    \item Dynamic Interests $\mathcal{U}_{\text{int}}$: Behavioral patterns derived from historical engagement, such as \textit{cycling enthusiast}, \textit{anime fan}, \textit{tech geek}, etc.
\end{itemize}

\item \textbf{Environmental Context} $\mathcal{E}$: Real-time multi-source contextual signals encompassing weather conditions, seasonal factors, and trending events. These signals provide temporal grounding for situational intent mining (e.g., \textit{rainy day}, \textit{winter season}, \textit{holiday sale}, etc.).
\end{enumerate}

\noindent
Together, these components form a rich hybrid context that captures both long-term preferences and real-time situational needs, which is formulated as:
\begin{equation*}
\mathcal{C} = \{\mathcal{B}, \mathcal{U}, \mathcal{E}\},
\end{equation*}
where behavioral entities in $\mathcal{B}$ are represented through atomic compression while user attributes and environmental signals retain natural language encoding to maintain semantic richness.

\paragraph{Intent Decomposition}
Given the hybrid context $\mathcal{C}$, the Global Planner performs deep reasoning to uncover latent user needs and decompose them into $K$ specialized personas $\{p_1, p_2, \ldots, p_K\}$, where each persona represents a distinct facet of user intent. The planner analyzes $\mathcal{C}$ through multi-dimensional reasoning by considering temporal trends, situational adaptation, and behavioral consistency to generate complementary personas that avoid cognitive overlap. Formally, the persona generation process can be expressed as:
\begin{equation}
\{p_1, p_2, \ldots, p_K\} = f_{\text{planner}}(\mathcal{C}),
\end{equation}
where $f_{\text{planner}}(\cdot)$ denotes the reasoning function. This design achieves two critical objectives: 

\begin{itemize}[topsep=0pt]
    \item Eliminating computational redundancy by performing intent decomposition once over the compressed context rather than having each expert independently process raw sequences.
    \item Ensuring cognitive coordination by explicitly orchestrating complementary reasoning perspectives, preventing experts from redundantly exploring overlapping semantic spaces.
\end{itemize}
The generated personas $\{p_1, \ldots, p_K\}$ are subsequently distributed to the Expert Ensemble (\S\ref{sec:experts}), where each expert agent adopts its assigned persona and conducts specialized item tag prediction.

\subsubsection{Distributed Experts}
\label{sec:experts}

Upon receiving specialized personas $\{p_1, \ldots, p_K\}$ from the Global Planner, the distributed expert ensemble executes parallel yet complementary item tag prediction tasks. Each expert agent operates under its assigned persona to generate a set of item tags that reflects a distinct facet of user intent. Formally, the expert prediction process can be expressed as:
\begin{equation}
\mathcal{T}_k = f_{\text{expert}}(p_k),
\end{equation}
where $f_{\text{expert}}(\cdot)$ denotes the expert reasoning function, and $\mathcal{T}_k = \{t_1^k, t_2^k, \ldots, t_{M_k}^k\}$ represents the set of predicted item tags for persona $p_k$, with $M_k$ denoting the generated tag count.

To enhance expert capabilities and satisfy multi-objective requirements in industrial recommendation scenarios, we further introduce a two-stage training strategy combining \textbf{Supervised Fine-Tuning (SFT)} and \textbf{Reinforcement Learning (RL)} optimization.

\paragraph{Stage 1: Supervised Fine-Tuning}
To establish foundational expert capabilities, we employ SFT on persona-aligned training samples. Given a persona $p_k$, we construct supervision signals from the user's subsequent interactions. Specifically, we leverage GPT-4 to identify which item categories from the user's next interactions semantically align with the persona's intent focus:
\begin{equation*}
\mathcal{C}_k^{\text{rel}} = \{c \in \mathcal{C}_{\text{next}} \mid f_{\text{GPT-4}}(c, p_k) = \text{True}\},
\end{equation*}
where $\mathcal{C}_{\text{next}}$ denotes all item categories from the user's subsequent interactions (held-out next behavior), and $f_{\text{GPT-4}}(\cdot)$ is a binary classifier that determines whether category $c$ is semantically relevant to persona $p_k$. To ensure sufficient supervision signals, we construct a fixed-size target label set $\mathcal{C}_k^{\text{target}}$ containing exactly 15 elements. If $|\mathcal{C}_k^{\text{rel}}| < 15$, we augment it with GPT-4-generated synthetic tags that follow the stylistic conventions of online category labels; if $|\mathcal{C}_k^{\text{rel}}| > 15$, we randomly sample 15 tags.

For each persona-target pair $(p_k, \mathcal{C}_k^{\text{target}})$, we train the expert model following the standard next-token-prediction training paradigm by minimizing the cross-entropy loss:
\begin{equation}
\mathcal{L}_{\text{SFT}}(\theta_{\text{expert}}) = -\mathbb{E}_{(p_k, \mathcal{C}_k^{\text{target}})} \left[ \log p_{\theta_{\text{expert}}}\left(\mathcal{C}_k^{\text{target}} \mid p_k \right) \right],
\end{equation}
where $p_{\theta_{\text{expert}}}(\cdot)$ denotes the expert model's output distribution. This supervised alignment ensures that expert agents learn to generate tags consistent with their assigned persona focus.

\textbf{Training Data Composition.} To balance domain-specific knowledge with general language capabilities, we mix persona-aligned recommendation data with a general-purpose corpus. The recommendation data comprises diverse contextual scenarios: pure behavioral patterns (32.17\%), trending events (6.97\%), weather-related contexts (1.19\%), and other contextual signals (7.36\%). To preserve the model's foundational linguistic and reasoning abilities, we incorporate general instruction-following data (52.31\%), ensuring that expert models maintain sufficient versatility and robustness. The complete training data composition is summarized in Table~\ref{tab:sft_data_mix}.
\begin{table}[t]
\centering
\captionsetup{justification=centering}
\caption{Data source distribution for supervised fine-tuning.}
\label{tab:sft_data_mix}
\begin{tabular}{lc}
\toprule
\textbf{Data Type} & \textbf{Proportion (\%)} \\
\midrule
\multicolumn{2}{l}{\textit{Recommendation Task}} \\
\quad $\bullet$\; Pure Behavior Patterns & 32.17 \\
\quad $\bullet$\; Trending Topics \& Events & 6.97 \\
\quad $\bullet$\; Weather-Related Contexts & 1.19 \\
\quad $\bullet$\; Other Situational Signals & 7.36 \\
\midrule
General Language Modeling & 52.31 \\
\bottomrule
\end{tabular}
\end{table}

\paragraph{Stage 2: Constrained Reinforcement Optimization}
Building upon the foundation established through supervised fine-tuning, we further introduce reinforcement learning optimization to enhance expert performance across multiple objectives (\textit{e.g.}, diversity, relevance, accuracy). Besides, to address the inherent conflicts in multi-reward optimization, we design a simple yet effective constrained reward shaping mechanism that balances competing objectives and improves overall performance.

\textbf{Policy Optimization Framework.}
For each input sample, we adopt the Group Relative Policy Optimization (GRPO) algorithm~\citep{shao2024deepseekmath,liu2024deepseek} to optimize the expert policy. Specifically, given an input context, we sample a group of $G$ outputs from the old policy $\pi_{\theta_{\text{old}}}$, and optimize the new policy $\pi_{\theta}$ by minimizing the following objective:
\begin{gather}
\label{eq:grpo}
\mathcal{L}_{\text{GRPO}}(\theta) = -\mathbb{E}_{(x,y) \sim \pi_{\theta_{\text{old}}}} \left[ \min\left( r(\theta) \hat{A}(x,y), \, \text{clip}\left(r(\theta), 1-\epsilon, 1+\epsilon\right) \hat{A}(x,y) \right) - \beta \cdot \mathbb{D}_{\text{KL}}\left(\pi_{\theta} \| \pi_{\text{ref}}\right) \right] , \\
\mathbb{D}_{\text{KL}}\left(\pi_{\theta} \| \pi_{\text{ref}}\right) = \frac{\pi_{\text{ref}}(y|x)}{\pi_{\theta}(y|x)} - \log \frac{\pi_{\text{ref}}(y|x)}{\pi_{\theta}(y|x)} - 1,
\end{gather}
where $r(\theta) = \frac{\pi_{\theta}(y|x)}{\pi_{\theta_{\text{old}}}(y|x)}$ denotes the importance sampling probability, $\hat{A}(x,y) = R(x,y) - \frac{1}{G}\sum_{i=1}^{G} R(x, y_i)$ is the group-normalized advantage, $R(x,y)$ is the reward function, $\epsilon$ is the clipping parameter, $\pi_{\text{ref}}$ is the reference policy (\textit{i.e.}, SFT base model), and $\beta$ controls the strength of the KL penalty. The KL divergence term prevents the policy from deviating too far from the reference model, ensuring training stability and mitigating reward hacking.

\textbf{Multi-Reward Modeling.}
To guide the model's learning direction effectively, we design a multi-objective reward function comprising four complementary components:

\begin{enumerate}[topsep=2pt, itemsep=2pt, label=(\roman*)]
\item \textbf{Accuracy Reward} $R_{\text{acc}}$: We encourage the expert to predict tags that align with online user behavior by measuring the recall against ground-truth interactions. Specifically, given predicted tags $\mathcal{T}_k = \{t_1, \ldots, t_M\}$ and interacted item categories $\mathcal{C}_{\text{gt}} = \{c_1, \ldots, c_N\}$, the reward is defined as:
\begin{equation*}
R_{\text{acc}} = \frac{1}{|\mathcal{C}_{\text{gt}}|} \sum_{c \in \mathcal{C}_{\text{gt}}} \mathbb{I}\left[ c \in f_{\text{tag2cat}}(\mathcal{T}_k) \right],
\end{equation*}
where $f_{\text{tag2cat}}(\cdot)$ maps predicted tags to item categories, and $\mathbb{I}[\cdot]$ is the indicator function. This metric quantifies how well the predicted tags cover the user's actual interests.

\item \textbf{Alignment Reward} $R_{\text{align}}$: To ensure that predicted tags align with human quality standards and the assigned persona's intent, we introduce an alignment reward based on human preference learning. Specifically, we train a dedicated reward model $f_{\text{RM}}(\cdot)$ using preference pairs constructed from RecGPT-V1's quality criteria (detailed in \S\ref{sec:judge_as_reward}). For each predicted tag $t_i \in \mathcal{T}_k$, we evaluate its alignment score with respect to persona $p_k$:
\begin{equation*}
R_{\text{align}} = \frac{1}{M_k} \sum_{i=1}^{M_k} f_{\text{RM}}(t_i, p_k),
\end{equation*}
where $f_{\text{RM}}(\cdot)$ is trained on positive and negative preference pairs labeled according to established quality standards, capturing both semantic relevance to the persona and human-judged output quality. The final alignment reward is the average score across all predicted tags, where higher values indicate better alignment with human expectations for the given persona.

\item \textbf{Diversity Reward} $R_{\text{div}}$: To encourage experts to explore diverse user interests within their assigned personas, we design a diversity reward that measures the semantic richness of predicted tags. Specifically, we encode tags using the BGE embedding model~\citep{bge_embedding} and compute the average cosine distance among tag representations:
\begin{equation*}
R_{\text{div}} = 1 - \frac{2}{M_k(M_k-1)} \sum_{i=1}^{M_k-1} \sum_{j=i+1}^{M_k} \frac{\mathbf{e}_i \cdot \mathbf{e}_j}{\|\mathbf{e}_i\| \|\mathbf{e}_j\|},
\end{equation*}
where $\mathbf{e}_i = f_{\text{BGE}}(t_i)$ denotes the embedding of tag $t_i$. Higher diversity scores encourage broader intent coverage without redundant predictions.

\item \textbf{Length Reward} $R_{\text{len}}$: To promote appropriate tag lengths that balance informativeness and retrieval effectiveness, we design a length-based reward. For each predicted tag $t$ with word number $l$, the reward is defined as:
\begin{equation*}
R_{\text{len}}(t) = \begin{cases}
1.0, & \text{if } 6 \leq l \leq 11, \\
0.5, & \text{if } 4 \leq l < 6 \text{ or } 11 < l \leq 13, \\
0.0, & \text{otherwise}.
\end{cases}
\end{equation*}
The overall length reward is the average across all predicted tags: $R_{\text{len}} = \frac{1}{M} \sum_{i=1}^{M} R_{\text{len}}(t_i)$, which avoids overly short tags that lack expressiveness and too long tags that hinder retrieval diversity.
\end{enumerate}

\begin{figure}[t]
    \centering
    \includegraphics[width=0.8\linewidth]{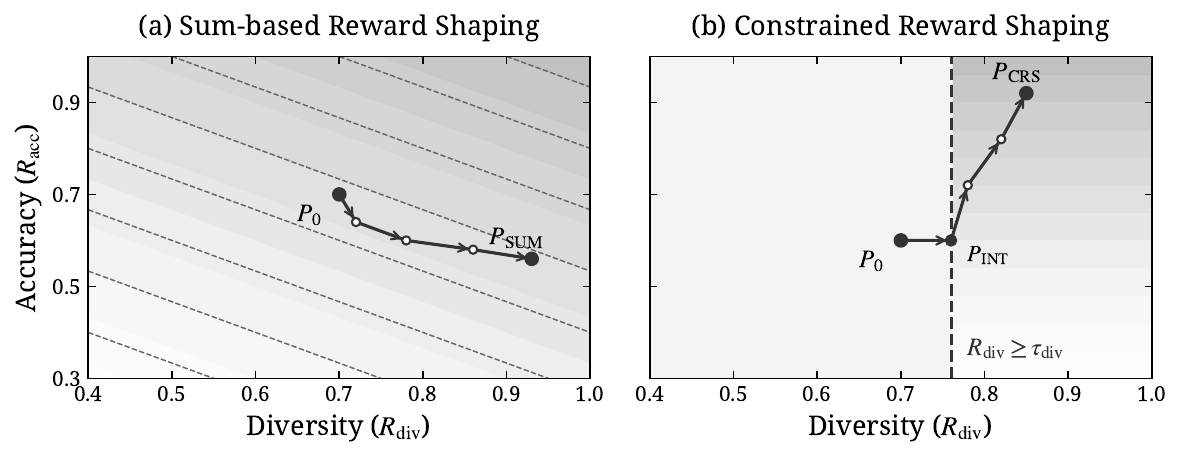}
    \caption{Comparison of reward shaping strategies. \textbf{(a)} Sum-based aggregation suffers from multi-reward conflicts. \textbf{(b)} Our constrained reward shaping treats secondary rewards (\textit{e.g.}, diversity) as conditional constraints, enabling stable optimization of the primary reward (\textit{i.e.}, accuracy).}
    \label{fig:reward_hacking}
\end{figure}

\textbf{Constrained Reward Shaping.}
Unlike conventional multi-objective reinforcement learning methods that directly sum individual rewards (denoted as \textbf{SUM}), we observe that such naive aggregation often leads to severe multi-reward conflicts. As illustrated in Figure~\ref{fig:reward_hacking}(a), the weighted-sum strategy mixes conflicting gradients across different reward dimensions, causing the optimization trajectory to drift toward suboptimal solutions (from $P_0$ to $P_{\text{SUM}}$) where simpler objectives (\textit{e.g.}, diversity) dominate at the expense of more critical objectives (\textit{e.g.}, accuracy).

To mitigate these conflicts, we propose a \textbf{Constrained Reward Shaping (CRS)} mechanism that treats certain rewards as hard constraints to guide the optimization of the primary accuracy objective. As shown in Figure~\ref{fig:reward_hacking}(b), our approach enforces a two-stage optimization process: the model first satisfies secondary constraints (moving from $P_0$ to $P_{\text{INT}}$ by crossing the feasibility boundary), and only then begins optimizing the primary accuracy reward (progressing from $P_{\text{INT}}$ to $P_{\text{CRS}}$). This design avoids gradient interference by decoupling constraint satisfaction from objective optimization. Formally, we define the composite reward as a product of conditional indicators:
\begin{equation}
R_{\text{total}} = R_{\text{acc}} \cdot \mathbb{I}[R_{\text{align}} \geq \tau_{\text{align}}] \cdot \mathbb{I}[R_{\text{div}} \geq \tau_{\text{div}}] \cdot \mathbb{I}[R_{\text{len}} \geq \tau_{\text{len}}],
\end{equation}
where $\mathbb{I}[\cdot]$ denotes the indicator function, and $\tau_{\text{align}}, \tau_{\text{div}}, \tau_{\text{len}}$ are predefined thresholds for alignment, diversity, and length rewards, respectively. This multiplicative formulation ensures that the accuracy reward is propagated only when all secondary objectives meet their minimum requirements. If any constraint is violated (\textit{i.e.}, any indicator returns 0), the total reward becomes zero, effectively mitigating conflicting gradient signals.

\begin{figure}[htbp]
    \centering
    \includegraphics[width=\linewidth]{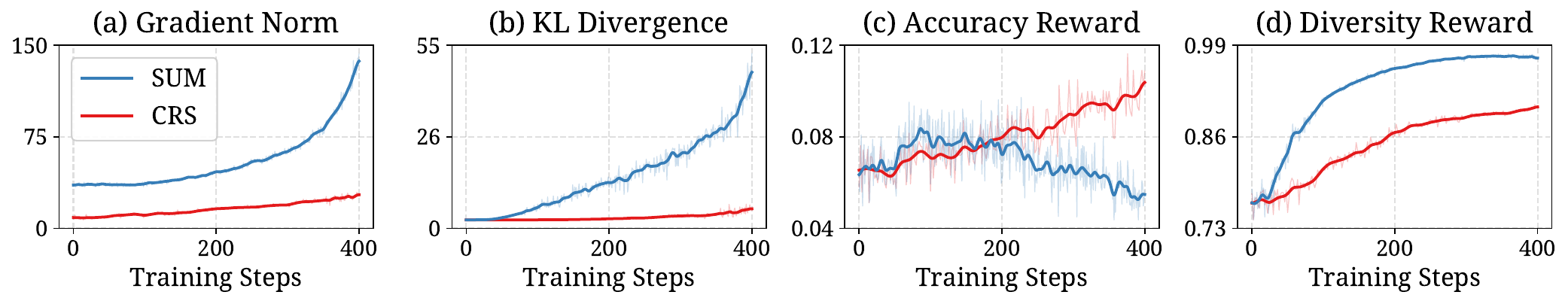}
    \caption{Training dynamics comparison between sum-based and constrained reward shaping. \textbf{(a)} Gradient norm. \textbf{(b)} KL divergence from reference model. \textbf{(c)} Accuracy reward. \textbf{(d)} Diversity reward. CRS maintains stable optimization across all metrics, while SUM suffers from multi-reward conflicts.}
    \label{fig:reward}
\end{figure}

As demonstrated in Figure~\ref{fig:reward}, our \textit{CRS} method exhibits superior training dynamics compared to the \textit{SUM} baseline. Figures~\ref{fig:reward}(a)-(b) show that \textit{CRS} maintains significantly lower gradient norms and KL divergence relative to the reference model, indicating improved training stability and reduced overfitting risk. Moreover, Figures~\ref{fig:reward}(c)-(d) reveal the fundamental limitation of additive aggregation: in later training stages, the optimization becomes dominated by simpler objectives (\textit{e.g.}, diversity), causing accuracy to degrade substantially due to gradient interference. In stark contrast, our \textit{CRS} method maintains simultaneous positive optimization across all objectives throughout training by decoupling constraint satisfaction from primary objective optimization, effectively mitigating multi-reward conflicts while preserving long-term optimization stability.

\begin{table}[h]
\centering
\caption{Tag prediction accuracy comparison across different training strategies. Both RecGPT-V1 and RecGPT-V2 variants are built upon Qwen-14B as the base model.}
\label{tab:expert_performance}
\begin{tabular}{lccccc}
\toprule
\multirow{2}{*}{\textbf{Metric}} & \multirow{2}{*}{\textbf{RecGPT-V1}} & \multicolumn{4}{c}{\textbf{RecGPT-V2}} \\
\cmidrule(lr){3-6}
 &  & \textbf{Base} & \textbf{SFT} & \textbf{GRPO (SUM)} & \textbf{GRPO (CRS)} \\
\midrule
HR@30 & 26.29\% & 23.08\% & 29.20\% & 27.38\% & \textbf{32.60\%} \\
\bottomrule
\end{tabular}
\end{table}

\paragraph{Experimental Evaluation}
Following RecGPT-V1, we adopt Hit Rate at top-30 predictions (HR@30) as the main evaluation metric, which measures whether the predicted item tags, after being mapped to item categories via a pre-trained Tag-to-Cate model, successfully match the user's actual interaction categories. Table~\ref{tab:expert_performance} presents the performance comparison across different training strategies.

The \textbf{Base} model underperforms RecGPT-V1 by 3.21\%, validating the necessity of domain adaptation. \textbf{SFT} substantially improves over both Base and RecGPT-V1 by 6.12\% and 2.91\% respectively, demonstrating that persona-aligned supervision effectively grounds expert reasoning. Comparing GRPO variants, \textbf{GRPO (SUM)} shows degraded performance relative to SFT, indicating that naive reward summation introduces gradient conflicts. In contrast, \textbf{GRPO (CRS)} achieves the highest HR@30 of 32.60\%, outperforming SFT by 3.40\% and RecGPT-V1 by 6.31\%, validating that treating secondary objectives as hard constraints enables stable reinforcement learning optimization.

\subsubsection{Decision Arbiter}
\label{sec:arbiter}

After the distributed expert ensemble generates complementary tag predictions $\{\mathcal{T}_1, \mathcal{T}_2, \ldots, \mathcal{T}_K\}$, the Decision Arbiter performs final candidate selection to produce a refined set of item tags for downstream retrieval. Given the aggregated tag pool $\mathcal{T}_{\text{all}} = \bigcup_{k=1}^{K} \mathcal{T}_k$ from all expert agents, the arbiter identifies the most promising tags that align with the user's real-time behavioral signals.

Specifically, the arbiter leverages the hybrid context $\mathcal{C} = \{\mathcal{B}, \mathcal{U}, \mathcal{E}\}$ to holistically evaluate all candidate tags in $\mathcal{T}_{\text{all}}$ across multiple quality dimensions (detailed criteria are provided in Appendix~\ref{appendix:eval_dimensions}). Rather than scoring tags individually, the arbiter performs joint reasoning over the entire candidate pool to identify the top-$N$ tags that collectively maximize behavioral relevance, profile consistency, content specificity, and validity:
\begin{equation*}
\mathcal{T}_{\text{final}} = f_{\text{arbiter}}(\mathcal{T}_{\text{all}}, \mathcal{C}).
\end{equation*}
This joint evaluation process enables the arbiter to consider inter-tag complementarity and avoid redundancy, effectively consolidating distributed expert outputs into a cohesive recommendation strategy that balances exploration breadth with focused user personalization.

\paragraph{Online Item Recommendation} 
After obtaining the refined tags, we further perform online item recommendation through multi-interest user encoding and traffic allocation optimization.

\textbf{Multi-Interest User Encoding.} Building upon RecGPT-V1's user-item-tag three-tower architecture, we extend the user encoder to capture multiple interest facets. Following the Poly-Encoder~\citep{humeau2019poly}, we introduce $K$ learnable context codes that aggregate user behavioral embeddings into multiple interest vectors $\{\mathbf{u}_1, \ldots, \mathbf{u}_K\}$ via attention mechanisms, where each vector represents a distinct aspect of user preferences. During online serving, the refined tags $\mathcal{T}_{\text{final}}$ are first encoded through the tag tower to obtain tag representations, which are then matched against items via the item tower. The multi-interest user representations are scored against candidate items through dot-product similarity, enabling fine-grained matching across diverse user intents.

\textbf{Traffic Allocation via Quadratic Programming.} To balance exploration (\textit{i.e.} cognitive channel) and exploitation (\textit{i.e.} existing utility channel) under limited exposure budgets, we formulate traffic allocation as a quadratic programming problem. This optimization framework dynamically adjusts the proportion of cognitive retrieval items in the recommendation slate, maximizing overall system revenue while ensuring that exploratory recommendations enhance long-term user engagement without compromising short-term business metrics. The detailed solution is provided in Appendix~\ref{appendix:implementation_details}.

\section{Dynamic Explanation Generation}
\label{sec:explanation}

Following RecGPT-V1, RecGPT-V2 retains the explanation generation module, providing personalized explanations to enhance user engagement with exposed items. However, extended online deployment reveals three critical deficiencies: \textbf{(1)} \textit{Low Information Density}—explanations frequently repeat generic phrases without conveying substantive insights; \textbf{(2)} \textit{Weak Temporal Adaptation}—failure to respond to seasonal trends, current events, or contextual signals; and \textbf{(3)} \textit{Homogenized Expression}—monotonous stylistic outputs that undermine user engagement. We attribute these deficiencies to two fundamental limitations: static prompt templates that constrain generative flexibility, and incomplete evaluation frameworks that neglect critical quality dimensions.

To address these challenges, this section introduces two key innovations in RecGPT-V2: \textbf{\textit{Meta-Prompting}} for dynamic explanation generation (\S\ref{sec:meta_prompting}), which synthesizes contextually adaptive prompt templates to enable diverse and situational explanations, and \textbf{\textit{preference-aware reinforcement learning}} (\S\ref{sec:exp_rl}), which optimizes generation quality through human-aligned multi-reward modeling. Together, these mechanisms transform explanation generation from template-based instantiation to dynamic reasoning, significantly enhancing user engagement and satisfaction.

\subsection{Meta-Prompting}
\label{sec:meta_prompting}

Unlike RecGPT-V1's direct one-step explanation generation from fixed templates, following current mainstream advances in prompt engineering~\citep{suzgun2024meta,zhang2023meta}, we introduce a \textbf{Meta-Prompting} framework that decomposes the generation process into two stages: \textit{style synthesis} followed by \textit{style-conditioned explanation generation}. This hierarchical design unlocks the model's creative capacity by first generating diverse, contextually adaptive stylistic guidelines, and then producing explanations that conform to these dynamic specifications.

\paragraph{Expanded Evaluation Dimensions}
We first extend RecGPT-V1's evaluation framework from four dimensions (Relevance, Factuality, Clarity, Safety) to seven dimensions by incorporating three additional criteria for better user experience: \textbf{(i)} \textit{Timeliness}, measuring alignment with current trends, seasonal contexts, or time-sensitive events; \textbf{(ii)} \textit{Informativeness}, quantifying the substantive insights conveyed beyond generic descriptions; and \textbf{(iii)} \textit{Attractiveness}, assessing the emotional appeal and persuasive power of the explanation. These expanded dimensions provide a more holistic assessment of explanation quality, guiding both meta-prompt generation and subsequent evaluation.

\paragraph{Two-Stage Generation Framework}
Given user interests, item attributes, and contextual signals (\textit{e.g.}, seasonal trends), the meta-prompting framework operates as follows:

\definecolor{niceBlue}{RGB}{52,101,164}
\definecolor{niceOrange}{RGB}{208,120,32}

\textbf{Stage 1: Style Synthesis.} The model first generates a stylistic guideline $g$ that specifies the desired tone, rhetorical devices, target audience, and emotional resonance. For example, given a children's toy item and a parent user profile during holiday seasons, the meta-prompt might produce:
``\textcolor{niceOrange}{\textit{Compose a playful, lighthearted, and visually evocative short caption that resonates with parents. Use a naive or gentle tone to create emotional connection.}}''

Formally, the style synthesis process can be expressed as:
\begin{equation}
g = f_{\text{meta}}(\mathcal{U}, \mathcal{I}, \mathcal{S}),
\end{equation}
where $\mathcal{U}$ denotes user interest, $\mathcal{I}$ represents item attributes, $\mathcal{S}$ encapsulates situational signals.

\textbf{Stage 2: Style-Conditioned Explanation Generation.} Conditioned on the style guideline $g$, the model generates the final explanation $e$ that adheres to the specified stylistic constraints:
\begin{equation}
e = f_{\text{exp}}(g, \mathcal{U}, \mathcal{I}, \mathcal{S}),
\end{equation}
where $f_{\text{exp}}(\cdot)$ denotes the explanation generation function. For instance, following the style guideline above, the model might produce:
``\textcolor{niceBlue}{\textit{Spins like a blue butterfly in the air.}}''
This two-stage decomposition provides a flexible framework that unleashes the model's imagination by enabling role-playing across diverse stylistic personas, delivering novel and contextually adaptive explanations to users.


\subsection{Preference-Aware Reinforcement Learning}
\label{sec:exp_rl}

Building upon RecGPT-V1's supervised fine-tuning foundation, we introduce constrained reinforcement learning to further enhance explanation quality, following the optimization framework in \S\ref{sec:experts}. We design a hybrid reward framework combining rule-based diversity rewards and model-based alignment rewards, unified under the Constrained Reward Shaping (CRS) mechanism.

\textbf{Policy Optimization Framework.}
Similar to \S\ref{sec:experts}, we adopt the GRPO algorithm to optimize the explanation generation policy. The optimization objective remains identical to Equation~\eqref{eq:grpo}, with the reward function replaced by the explanation-specific composite reward defined below.

\textbf{Hybrid Reward Modeling.}
To guide explanation generation across multiple quality dimensions, we design a hybrid reward function comprising two complementary components:

\begin{enumerate}[topsep=2pt, itemsep=2pt, label=(\roman*)]
\item \textbf{Rule-Based Diversity Reward} $R_{\text{div}}$: To encourage varied linguistic expressions and avoid repetitive patterns, we design an IDF-inspired diversity reward. We maintain a memory buffer $\mathcal{M}$ of size 160 that stores recent generated explanations in tokenized form, updated in FIFO manner. For each newly generated explanation $e = \{w_1, w_2, \ldots, w_L\}$, the diversity score is:
\begin{equation*}
R_{\text{div}} = \frac{1}{L} \sum_{i=1}^{L} \log \frac{|\mathcal{M}|}{\left|\{e' \in \mathcal{M} : w_i \in e'\}\right| + 1},
\end{equation*}
where $|\mathcal{M}|$ is the buffer size, and $|\{e' \in \mathcal{M} : w_i \in e'\}|$ counts stored explanations containing token $w_i$. The logarithmic term assigns higher rewards to rare tokens that enhance lexical diversity, with $+1$ smoothing preventing division by zero.

\item \textbf{Model-Based Alignment Reward} $R_{\text{align}}$: To capture subjective quality dimensions (\textit{e.g.}, informativeness), we train a reward model $f_{\text{RM}}(\cdot)$ on preference data using listwise comparisons (detailed in \S\ref{sec:judge_as_reward}). Given a generated explanation $e$, the alignment reward is computed as:
\begin{equation*}
R_{\text{align}} = f_{\text{RM}}(e, \mathcal{U}, \mathcal{I}, \mathcal{S}).
\end{equation*}
\end{enumerate}

\textbf{Constrained Reward Shaping.}
Consistent with \S\ref{sec:experts}, we adopt CRS to mitigate multi-reward conflicts. Here, explanation generation prioritizes \textit{human preference alignment} as the main reward, with diversity as a secondary constraint. Therefore, the total reward is formulated as:
\begin{equation}
R_{\text{total}} = R_{\text{align}} \cdot \mathbb{I}\left[R_{\text{div}} \geq \tau_{\text{div}}\right],
\end{equation}
where $\tau_{\text{div}}$ is the diversity threshold. By treating diversity as a gating condition, CRS eliminates gradient interference and enables stable optimization toward human-aligned, diverse explanations.

\paragraph{Experimental Evaluation}

We evaluate explanation generation quality through both diversity and human evaluation.

\textbf{Diversity.} We measure explanation diversity by computing pairwise dissimilarity within explanation sets generated for each item. Specifically, for each item $i$ with generated explanation set $\{e_1^i, e_2^i, \ldots, e_K^i\}$, we compute the diversity score as:
\begin{equation*}
\text{Diversity}_i = 1 - \frac{2}{K(K-1)} \sum_{j=1}^{K-1} \sum_{k=j+1}^{K} \text{ROUGE-L}(e_j^i, e_k^i),
\end{equation*}
where ROUGE-L measures the longest common subsequence similarity between explanation pairs. Higher scores indicate greater lexical diversity across generated explanations.

\textbf{Quality.} We conduct human annotation to assess explanation quality across the seven evaluation dimensions (\textit{cf.} Table~\ref{tab:explanation_eval_dimensions}). Annotators label explanations as high-quality if they satisfy all criteria.

\begin{wraptable}{r}{0.48\textwidth}
\vspace{-14pt}
\centering
\caption{Explanation performance comparison.}
\vspace{-2pt}
\label{tab:exp_performance}
\begin{tabular}{lcc}
\toprule
\textbf{Method} & \textbf{Diversity} & \textbf{Quality (\%)} \\
\midrule
RecGPT-V1 & 0.631 & 36.03 \\
RecGPT-V2 & \textbf{0.677} & \textbf{40.73} \\
\bottomrule
\end{tabular}
\vspace{-8pt}
\end{wraptable}

Table~\ref{tab:exp_performance} shows that RecGPT-V2 achieves substantial improvements: diversity increases by \textcolor{red}{\textbf{7.30\%}} and quality acceptance rate improves by \textcolor{red}{\textbf{13.04\%}}. These gains validate the effectiveness of meta-prompting and preference-aware reinforcement learning. The meta-prompting mechanism enables contextually adaptive style synthesis, while CRS optimization ensures that both diversity and quality improve simultaneously by eliminating gradient interference.
\section{Agentic Judge Framework}
\label{sec:judge}

To evaluate recommendation generation tasks, RecGPT-V1 introduced an \textit{LLM-as-a-Judge} method to reduce the inefficiency and high cost of human annotation. However, this outcome-focused approach directly predicts quality scores without decomposing the evaluation into intermediate reasoning steps, limiting its ability to capture nuanced quality distinctions across multiple dimensions. This collapsed evaluation paradigm overlooks the multi-step deliberation process that human evaluators employ, resulting in suboptimal alignment with human judgment standards.

To further enhance evaluation quality, RecGPT-V2 introduces a novel evaluation paradigm comprising two innovations: \textbf{\textit{Agent-as-a-Judge}} (\S\ref{sec:agentic_judge}), which decomposes complicated quality assessment into Multi-Dimension progressive reasoning, and \textbf{\textit{Judge-as-a-Reward}} (\S\ref{sec:judge_as_reward}), which distills agent judgments into dense reward signals for reinforcement learning optimization. 
Together, these designs establish a self-reinforcing \textbf{Flywheel Effect}: the policy model generates diverse outputs, agentic evaluation provides Multi-Dimension quality feedback, and reward distillation converts assessments into optimization signals for reinforcement learning.

\subsection{Agent-as-a-Judge}
\label{sec:agentic_judge}

\begin{figure}
    \centering
    \includegraphics[width=0.9\linewidth]{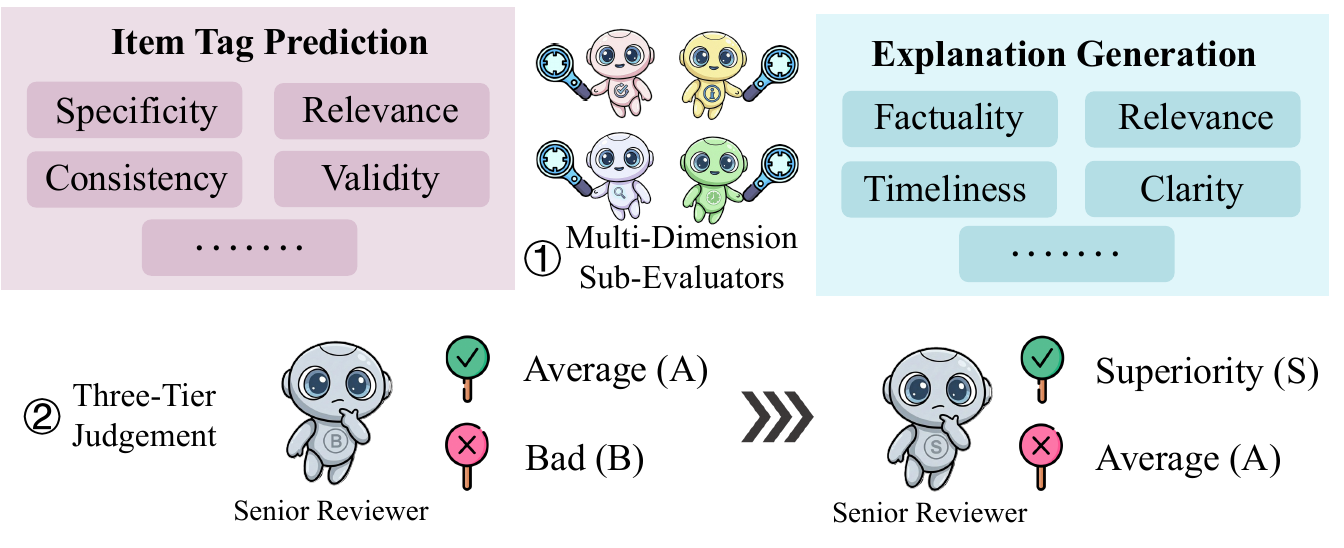}
    \caption{Agent-as-a-Judge framework mimicking human process-oriented fine-grained evaluation. Multi-dimension sub-evaluators independently assess specialized quality dimensions, and Senior Reviewer aggregates feedback into three-tier judgments (Superior/Average/Bad).}
    \label{fig:agent_as_a_judge}
\end{figure}

Unlike RecGPT-V1's end-to-end LLM-as-a-Judge evaluation, motivated by recent works~\citep{zhuge2024agent,gou2025mind2web,zhang2025sentient}, we introduce an \textbf{Agent-as-a-Judge} framework that mirrors human cognitive evaluation through hierarchical multi-agent reasoning. This design decomposes holistic quality assessment into fine-grained, dimension-specific sub-evaluators followed by multi-level review, enabling more accurate and interpretable quality judgments.

\paragraph{Multi-Dimension Sub-Evaluators}
For recommendation generation tasks involving multiple evaluation dimensions (\textit{cf.} Appendix~\ref{appendix:eval_dimensions}), we instantiate a specialized sub-evaluator for each dimension. Each sub-evaluator $\mathcal{E}_i$ assesses the generated content $y$ along its assigned dimension $d_i$:
\begin{equation*}
s_i = \mathcal{E}_i(y, d_i),
\end{equation*}
where $s_i$ represents the dimension-specific evaluation result. This decomposition transforms the complex multi-objective evaluation into manageable single-objective sub-tasks, enabling each evaluator to specialize in capturing nuanced quality aspects.

\paragraph{Three-Tier Judgment}
To derive a final overall quality, we introduce a \textbf{Senior Reviewer Agent} that aggregates the outputs $\{s_1, \ldots, s_D\}$ from all sub-evaluators. The Senior Reviewer produces the final decision using a three-tier S-A-B scheme:
\begin{itemize}[topsep=2pt, itemsep=2pt]
    \item \textbf{Superior (S)}: Output excels across all or most dimensions.
    \item \textbf{Average (A)}: Output meets minimum standards across dimensions.
    \item \textbf{Bad (B)}: Output fails to satisfy basic requirements in at least one critical dimension.
\end{itemize}

The aggregation procedure operates through a two-stage decision process:
\begin{enumerate}[topsep=2pt, itemsep=2pt, label=(\alph*)]
    \item \textbf{Defect Detection}: If any dimension receives a negative or unsatisfactory signal, the overall result is classified as \textbf{B}.
    \item \textbf{Excellence Elevation}: If no critical defects are detected, the Senior Reviewer further distinguishes between \textbf{S} and \textbf{A} based on the proportion or pattern of positive feedback among all dimensions, using a threshold $\tau$ to control the stringency for high-quality classification.
\end{enumerate}

\paragraph{Model Adaptation through Supervised Fine-Tuning}
To adapt the evaluation agents to domain-specific quality standards, we construct a training corpus combining model-generated samples and outputs from powerful LLMs (\textit{e.g.}, DeepSeek-R1~\citep{guo2025deepseek}, Qwen3-235B~\citep{yang2025qwen3technicalreport}). To ensure sufficient coverage of Bad-quality samples, we employ a hybrid annotation strategy: \textbf{(1)} for dimensions such as relevance, we automatically construct training samples through in-batch shuffling by randomly pairing outputs with mismatched user contexts; \textbf{(2)} for dimensions requiring nuanced judgment, human annotators provide labels across all evaluation dimensions, including both dimension-specific assessments $\{s_1, \ldots, s_D\}$ and holistic S-A-B judgments. We fine-tune a lightweight Qwen3-32B-Instruct model on this mixed training data using a SFT training paradigm.

\begin{table}[h]
\centering
\caption{Human-Judge agreement comparison on Superior (S) quality identification between LLM-as-a-Judge (RecGPT-V1) and Agent-as-a-Judge (RecGPT-V2) across three models, where human annotations serve as the ground truth. The best results are highlighted in \textbf{bold}.}
\label{tab:judge_comparison}
\begin{tabular}{clcccc}
\toprule
\multirow{2}{*}{\textbf{Task}} & \multicolumn{1}{c}{\multirow{2}{*}{\textbf{Model}}} & \multicolumn{2}{c}{\textbf{Accuracy}} & \multicolumn{2}{c}{\textbf{F1}} \\
\cmidrule(lr){3-4} \cmidrule(lr){5-6}
 & & V1 & V2 & V1 & V2 \\
\midrule
\multirow{3}{*}{\makecell[c]{Item Tag\\Prediction}} 
 & GPT5-mini  & 0.7694 & \textbf{0.7704} & 0.7499 & \textbf{0.7535} \\
 & Qwen3-Base & 0.7844 & \textbf{0.7864} & 0.7991 & \textbf{0.8051} \\
 & Qwen3-SFT  & 0.8210 & \textbf{0.8248} & 0.8095 & \textbf{0.8228} \\
\midrule
\multirow{3}{*}{\makecell[c]{Explanation\\Generation}} 
 & GPT5-mini  & 0.4481 & \textbf{0.4548} & \textbf{0.5673} & 0.5424 \\
 & Qwen3-Base & \textbf{0.3423} & 0.2764 & 0.0898 & \textbf{0.0904} \\
 & Qwen3-SFT  & 0.6885 & \textbf{0.7006} & 0.6787 & \textbf{0.7307} \\
\bottomrule
\end{tabular}
\end{table}

\paragraph{Experimental Evaluation}
Table~\ref{tab:judge_comparison} presents the human-judge agreement comparison between LLM-as-a-Judge (RecGPT-V1) and Agent-as-a-Judge (RecGPT-V2) on Superior (S) quality identification, where human annotations serve as the ground truth. RecGPT-V2 demonstrates consistent improvements across most experimental configurations. For item tag prediction, Agent-as-a-Judge achieves higher accuracy across all three models (+0.10pp, +0.20pp, +0.38pp) with corresponding F1 improvements (+0.36pp, +0.60pp, +1.33pp), indicating enhanced precision-recall balance. For explanation generation, RecGPT-V2 maintains superior performance in GPT5-mini (+0.67pp accuracy) and Qwen3-SFT (+1.21pp accuracy, +5.20pp F1), with the latter showing the most substantial F1 gain. These results validate that decomposing holistic quality assessment into dimension-specific sub-evaluations followed by senior reviewer aggregation enhances human-AI alignment, achieving more reliable quality identification while maintaining computational efficiency for industrial deployment.

\subsection{Judge-as-a-Reward}
\label{sec:judge_as_reward}

While Agent-as-a-Judge provides accurate quality assessment, directly applying it for reinforcement learning optimization faces two challenges: \textbf{(1)} discrete classification labels lack the granularity needed for fine-grained policy gradient estimation, and \textbf{(2)} the multi-step evaluation incurs high computational overhead during online RL training. To address these issues, we introduce \textbf{Judge-as-a-Reward}, a distillation framework that transfers agent evaluation capabilities into lightweight reward models for providing dense optimization signals.

\paragraph{Reward Model Architecture}
We initialize the reward model from the Agent Judge checkpoint, inheriting its learned evaluation knowledge. The key architectural modification replaces the language modeling head with a scalar value head:
\begin{equation*}
r = f_{\text{RM}}(y, \mathcal{U}, \mathcal{I}, \mathcal{S}),
\end{equation*}
where $f_{\text{RM}}(\cdot)$ denotes the reward model, and $r \in \mathbb{R}$ represents the predicted reward score conditioned on generated content $y$, user interests $\mathcal{U}$, item attributes $\mathcal{I}$, and situational signals $\mathcal{S}$. The value head applies a sigmoid activation to bound outputs into $[0,1]$, facilitating stable gradient flow.

\paragraph{Reward Model Training via Listwise Learning-to-Rank}
To preserve fine-grained quality distinctions from the Senior Reviewer's three-tier labels, we adopt a listwise learning-to-rank approach. For each training batch, samples are grouped by their assigned quality level (\textbf{S}, \textbf{A}, \textbf{B}). For any quality level $g$, samples at level $g$ serve as positive instances, while all samples at lower levels constitute the negative set. The reward model is trained to assign higher scores to higher-quality samples using the following unified contrastive loss formulation:
\begin{equation}
\mathcal{L}_{\text{RM}} = -\sum_{g \in \{\text{S}, \text{A}\}} \sum_{y_g \in \mathcal{Y}_g} \log \frac{\exp(f_{\text{RM}}(y_g))}{\exp(f_{\text{RM}}(y_g)) + \sum_{g' < g} \sum_{y_{g'} \in \mathcal{Y}_{g'}} \exp(f_{\text{RM}}(y_{g'}))},
\end{equation}
where $g' < g$ denotes all quality levels lower than $g$ (\textit{e.g.}, for $g=\text{S}$, negatives include both \textbf{A} and \textbf{B}; for $g=\text{A}$, negatives include only \textbf{B}), and $\mathcal{Y}_g$ represents the set of samples at level $g$. This formulation implicitly captures all pairwise relationships (\textbf{S} vs \textbf{AB}, \textbf{A} vs \textbf{B}), enabling the reward model to learn the complete hierarchical preference ordering from annotated data.

\textit{Engineering Acceleration via Prefix Sharing.} To accelerate training, we exploit the observation that samples within each contrastive group share identical contextual prompts, differing only in generated content. By computing shared prefix representations once and reusing them across all candidates, we enable parallel inference and significantly reduce redundant computation.

\paragraph{Self-Improving Flywheel Effect}
The synergistic integration of Agent-as-a-Judge and Judge-as-a-Reward establishes a self-reinforcing optimization cycle that enables continuous quality improvement without recurring human annotation costs:

\begin{enumerate}[wide, topsep=2pt, itemsep=2pt, label=\textbf{Stage \arabic*:}]
    \item \textbf{Policy Generation.} The policy model explores the output space through supervised fine-tuning and reinforcement learning, generating diverse responses across varying quality levels.
    
    \item \textbf{Agentic Evaluation.} The Agent-as-a-Judge framework decomposes each generated sample into dimension-specific quality assessments, synthesizing these into holistic S-A-B tier judgments through the Senior Reviewer's deliberation process.

    \item \textbf{Reward Distillation.} The Judge-as-a-Reward model distills the discrete agent judgments into continuous, differentiable, and more informative reward signals by learning the underlying preference structure through listwise contrastive training.

    \item \textbf{Policy Optimization.} The distilled reward signals guide policy refinement via GRPO (\S\ref{sec:experts}), updating model parameters to maximize expected human-aligned preferences.
\end{enumerate}

This closed-loop architecture creates a \textbf{\textit{flywheel effect}}: as the policy generates higher-quality outputs, the agent evaluator accumulates richer training signals, which improve reward model calibration and enable more effective policy optimization. Critically, this cycle operates autonomously after initial human annotation, progressively aligning model behavior with human quality standards. The reward distillation ensures computational efficiency for rapid iteration, while Multi-Dimension evaluation guarantees quality improvements across all criteria rather than narrow metric optimization.

\begin{table}[h]
\centering
\caption{Performance comparison of reward model training strategies. \textbf{HR@30} denotes hit rate at top-30 for item tag prediction. \textbf{Quality} measures human-evaluated explanation superior rate.}
\label{tab:reward_model_comparison}
\begin{tabular}{lcc}
\toprule
\textbf{Method} & \textbf{HR@30 (Tag)} & \textbf{Quality (Explanation)} \\
\midrule
RecGPT-V1 & 26.29\% & 36.03\% \\
RecGPT-V2 (Point-wise RM) & 31.24\% & 37.64\% \\
RecGPT-V2 (List-wise RM) & \textbf{32.60\%} & \textbf{40.73\%} \\
\bottomrule
\end{tabular}
\end{table}

\paragraph{Experimental Evaluation}
Table~\ref{tab:reward_model_comparison} compares the impact of different reward model training strategies on reinforcement learning performance across item tag prediction (HR@30) and explanation generation (Quality). RecGPT-V2 with listwise reward modeling achieves obvious improvements over RecGPT-V1 (+24.1\% HR@30, +13.0\% Quality) and pointwise training (+4.4\% HR@30, +8.2\% Quality). The listwise learning-to-rank formulation captures the complete hierarchical preference ordering (\textbf{S} $\succ$ \textbf{A} $\succ$ \textbf{B}) by modeling all pairwise relationships simultaneously, enabling the reward model to provide more discriminative optimization signals that guide policy learning toward human-aligned quality standards. In contrast, pointwise training treats samples independently, losing the relative preference structure critical for effective policy gradient estimation.

\section{Experiments}
\label{sec:experiments}

To validate the effectiveness of RecGPT-V2 in practical industrial application, we conduct long-term online experiments on Taobao's platform. In the following sections, we detail the online A/B test performance and real-world case study to illustrate the advantages of our proposed system.

\subsection{Online A/B Test}
\label{sec:online_ab}

\paragraph{Experimental Setup}
We deploy RecGPT-V2 on Taobao's homepage ``Guess What You Like'' scenario, conducting a two-week online A/B test with the following configuration:

\begin{itemize}[topsep=2pt, itemsep=2pt]
    \item \textbf{Traffic Allocation}: Both experimental and control groups each receive 1\% of total platform traffic, ensuring statistically significant results while minimizing deployment risk.
    \item \textbf{Baseline Comparison}: RecGPT-V1 serves as the control group, allowing direct assessment of the improvements introduced in RecGPT-V2.
    \item \textbf{Evaluation Scenarios}: We separately evaluate performance in two distinct scenarios:
    \begin{itemize}[topsep=2pt, itemsep=2pt]
        \item \textit{Item Scenario}: Direct item recommendations displayed in grid layout.
        \item \textit{Feed Scenario}: Mixed-content recommendation stream in the main feed, including items, advertisements, live streams, and other content types.
    \end{itemize}
\end{itemize}

\paragraph{Evaluation Metrics}
To comprehensively assess system performance, we measure both short-term engagement and long-term retention metrics, defined as follows:

\textbf{Short-Term Metrics}:
\begin{itemize}[topsep=2pt, itemsep=2pt]
    \item \textit{IPV (Item Page Views)}: Number of item detail page visits, indicating user interest.
    \item \textit{CTR (Click-Through Rate)}: Ratio of clicks to impressions, measuring recommendation relevance.
    \item \textit{TV (Transaction Volume)}: Monetary value of completed purchases.
    \item \textit{GMV (Gross Merchandise Value)}: Total transaction value including orders and returns.
    \item \textit{ATC (Add-to-Cart)}: Number of items added to shopping cart, reflecting purchase intent.
\end{itemize}

\textbf{Long-Term Metrics}:
\begin{itemize}[topsep=2pt, itemsep=2pt]
    \item \textit{NER (Novelty Exposure Rate)}: Percentage of recommended items that users have not previously interacted with, measuring exploration effectiveness.
    \item \textit{LT-14 / LT-30}: User retention rates at 14-day and 30-day horizons, quantifying long-term engagement sustainability.
\end{itemize}

\begin{table}[h]
\centering
\caption{Online A/B test results comparing RecGPT-V2 against RecGPT-V1 baseline across item and feed scenarios. All metrics show relative percentage improvements (\% omitted).}
\label{tab:online_ab_results}
\begin{threeparttable}
\begin{tabular}{cccccccccc}
\toprule
\multirow{2}{*}{\textbf{Scenario}} & \multicolumn{5}{c}{\textbf{Short-Term Engagement}} & \multicolumn{3}{c}{\textbf{Long-Term Retention}} \\
\cmidrule(lr){2-6} \cmidrule(lr){7-9}
& \textbf{IPV} & \textbf{CTR} & \textbf{TV} & \textbf{GMV} & \textbf{ATC} & \textbf{NER} & \textbf{LT-14} & \textbf{LT-30} \\
\midrule
\textbf{Item} & +3.64 & +3.01 & +2.11 & +3.39 & +3.47 & +11.46 & -- & -- \\
\textbf{Feed} & +1.29 & +1.50 & +0.34 & +1.53 & +0.99 & +4.49 & +0.04 & +0.05 \\
\bottomrule
\end{tabular}%
\begin{tablenotes}
\small
\item Note: -- indicates metrics not applicable in the item scenario.
\end{tablenotes}
\end{threeparttable}
\end{table}

\paragraph{Results and Analysis}
Table~\ref{tab:online_ab_results} summarizes the online A/B test results. RecGPT-V2 consistently outperforms the RecGPT-V1 baseline across both scenarios and all metrics, demonstrating substantial improvements in user engagement and platform value.

$\blacklozenge$\; Across short-term engagement metrics, RecGPT-V2 achieves notable gains in the item scenario, with IPV, CTR, TV, GMV, and ATC improving by +3.26\%, +3.01\%, +2.11\%, +3.39\%, and +3.47\% respectively. These improvements suggest that the enhanced intent understanding translates directly to increased user interaction and transaction value. The feed scenario exhibits consistent positive trends, with CTR (+1.50\%) and GMV (+1.53\%) gains indicating improved recommendation relevance.

$\blacklozenge$\; The long-term retention metrics reveal particularly striking results. NER increases by +11.46\% in the item scenario and +4.49\% in the feed scenario, indicating substantially improved recommendation diversity and novelty. This finding validates our hypothesis that multi-agent coordination and environmental signal integration effectively mitigate filter bubble effects. While LT-14 and LT-30 improvements appear modest in absolute terms (+0.04\% and +0.05\%), these gains represent meaningful progress in sustained user retention, which is critical for platform health.

\subsection{Case Study}

\begin{figure}[h]
    \centering
    \captionsetup{justification=centering}
    \includegraphics[width=0.8\textwidth]{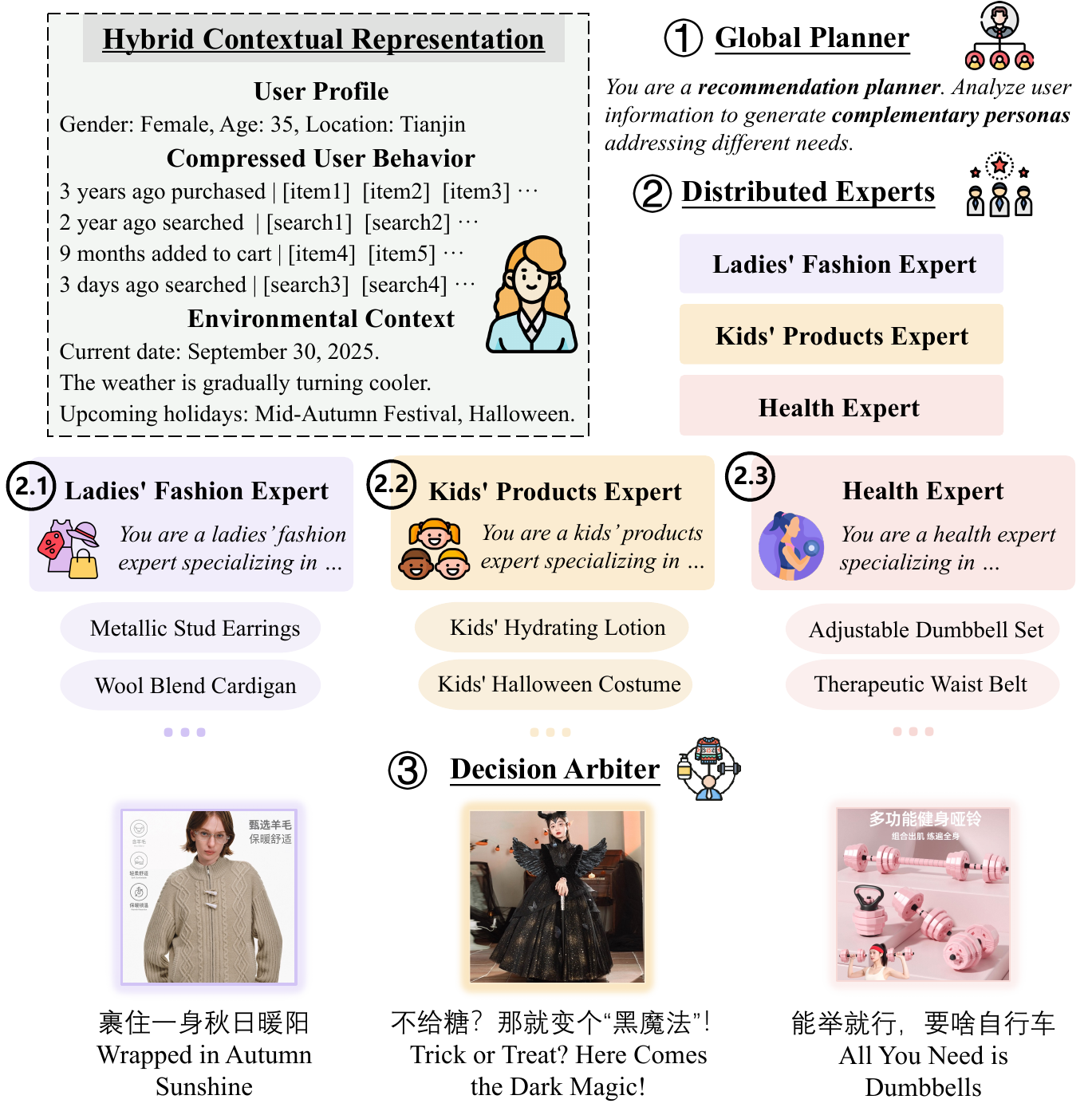}
    \caption{Case study.}
    \label{fig:case_study}
\end{figure}

Figure~\ref{fig:case_study} illustrates a real-world case demonstrating RecGPT-V2's strengths in dynamic intent understanding and context-aware recommendation generation. Given a user profile (35-year-old female, Tianjin) with compressed behavioral history, the system ingests real-time environmental signals including \textbf{cooling weather}, upcoming \textbf{Mid-Autumn Festival} and \textbf{Halloween}. The Global Planner decomposes these contextual signals into three complementary personas: \textit{Ladies' Fashion Expert}, \textit{Kids' Products Expert}, and \textit{Health Expert}. Each expert independently generates domain-specific item tags through specialized reasoning: the fashion expert predicts ``\textcolor[HTML]{D99058}{\textbf{Wool Blend Cardigan}}'' responding to the cooling weather; the kids expert generates both ``\textcolor[HTML]{D99058}{\textbf{Kids' Hydrating Lotion}}'' (addressing dry autumn climate) and ``\textcolor[HTML]{D99058}{\textbf{Kids' Halloween Costume}}'' (anticipating the upcoming holiday), demonstrating temporal adaptation; the health expert recommends ``\textcolor[HTML]{D99058}{\textbf{Adjustable Dumbbell Set}}'' aligning weather-driven wellness needs with historical fitness interests. The Decision Arbiter synthesizes expert predictions and selects three final items, each paired with contextually adaptive explanations generated by the meta-prompting framework, such as ``\textcolor{purple}{\textit{Wrapped in Autumn Sunshine}}'' (emphasizing seasonal comfort), ``\textcolor{purple}{\textit{Quench Your Little One's Skin}}'' (highlighting children's autumn skincare), and ``\textcolor{purple}{\textit{All You Need is Dumbbells}}'' (promoting accessible home fitness). This case validates RecGPT-V2's core capability: by integrating real-time environmental signals into hierarchical multi-agent reasoning, the system achieves both diverse intent coverage and precise situational adaptation, moving beyond static behavioral pattern matching toward dynamic, context-aware recommendation generation.
\section{Conclusion}
\label{sec:conclusion}

This paper presents RecGPT-V2, an agentic framework that advances LLM-powered recommender systems through agentic intent reasoning, meta-prompting for explanation generation, constrained reinforcement learning, and process-oriented Agent-as-a-Judge evaluation. By eliminating cognitive redundancy and optimizing computational efficiency, RecGPT-V2 reduces GPU consumption by 60\% while improving generation quality across both item tag prediction and explanation tasks. Large-scale deployment on Taobao demonstrates significant online gains (\textit{e.g.}, +3.40\% IPV, +4.68\% CTR, +4.05\% TV, +11.46\% NER), validating the practical viability of integrating LLM-based intent reasoning into industrial recommender systems at scale.
In future work, we aim to further explore how to end-to-end jointly optimize multi-agent collaboration with reinforcement learning techniques to enhance recommendation performance and user experience.

\addcontentsline{toc}{section}{References}
\bibliographystyle{abbrvnat}
\nobibliography*
\bibliography{reference}

\clearpage

\appendix
\section*{Appendix}
\section{Contributors}

\begin{multicols}{2}
\noindent
\textcolor[HTML]{5b0f08}{
\textbf{Core Contributors} \\
Chao Yi \\
Dian Chen \\
Gaoyang Guo \\
$\text{Jiakai Tang}^{\dagger}$ \\
Jian Wu \\
Jing Yu \\
Mao Zhang \\
Wen Chen \\
Wenjun Yang \\
Yujie Luo \\
Yuning Jiang \\
Zhujin Gao \\
}


\noindent
\textcolor[HTML]{030361}{
\textbf{Contributors} \\
Bo Zheng \\ 
Binbin Cao \\
Changfa Wu \\
Dixuan Wang \\ 
Han Wu \\ 
Haoyi Hu \\
Kewei Zhu \\
Lang Tian \\ 
Lin Yang \\ 
Qiqi Huang \\ 
Siqi Yang \\
Wenbo Su \\ 
Xiaoxiao He \\ 
Xin Tong \\ 
$\text{Xu Chen}^{\dagger}$ \\ 
Xunke Xi \\ 
Xiaowei Huang \\ 
Yaxuan Wu \\ 
Yeqiu Yang \\
Yi Hu \\ 
Yujin Yuan \\ 
Yuliang Yan \\ 
Zile Zhou \\ 
}
\end{multicols}

\noindent
$\dagger$ Renmin University of China

\noindent
The listing of authors is in alphabetical order based on their first names.

\newpage

\section{Evaluation Dimensions for Generation Tasks}
\label{appendix:eval_dimensions}

The detailed evaluation dimensions for item tag prediction and recommendation explanation generation tasks are listed in Table~\ref{tab:tag_eval_dimensions} and Table~\ref{tab:explanation_eval_dimensions}, respectively.

\begin{table}[h]
\centering
\captionsetup{justification=centering}
\caption{Evaluation dimensions for item tag prediction task.}
\label{tab:tag_eval_dimensions}
\begin{tabularx}{\textwidth}{lX}
\toprule
\multicolumn{1}{c}{\textbf{Dimension}} & \multicolumn{1}{c}{\textbf{Definition}} \\
\midrule
\textbf{Relevance} & Evaluates whether the tags are directly aligned with the user's associated interests. This criterion measures the model's capacity to genuinely understand and accurately predict user needs by assessing whether the tag matches the specified interest. \\
\addlinespace
\textbf{Consistency} & Assesses whether the item tag is generated with explicit reference to the user's profile information and historical behavioral data. This criterion focuses on whether the model's reasoning process incorporates authentic user context rather than fabricating or ignoring the given user information. \\
\addlinespace
\textbf{Specificity} & Evaluates tag specificity to avoid generic terms like ``fashion sports equipment'' that lead to imprecise product retrieval. \\
\addlinespace
\textbf{Validity} & Determines whether the predicted tags correspond to an actual existing product, preventing non-existent tag generation. \\
\bottomrule
\end{tabularx}
\end{table}

\begin{table}[h]
\centering
\caption{Evaluation dimensions for recommendation explanation generation task, where dimensions marked with $^*$ are newly introduced in RecGPT-V2.}
\label{tab:explanation_eval_dimensions}
\begin{tabularx}{\textwidth}{lX}
\toprule
\multicolumn{1}{c}{\textbf{Dimension}} & \multicolumn{1}{c}{\textbf{Definition}} \\
\midrule
\textbf{Relevance} & Alignment between the explanation and both the characteristics of the recommended item and the user's interests. \\
\addlinespace
\textbf{Factuality} & Accuracy of the explanation in reflecting the item's actual features. \\
\addlinespace
\textbf{Clarity} & Quality of text fluency, grammatical correctness, and stylistic expression. \\
\addlinespace
\textbf{Safety} & Absence of sensitive or personal information in the generated content. \\
\addlinespace
$\text{\textbf{Timeliness}}^*$ & Alignment with seasonal trends, current events, or temporal contexts. \\
\addlinespace
$\text{\textbf{Informativeness}}^*$ & Degree to which the explanation provides useful and detailed information about the item, enhancing user understanding. \\
\addlinespace
$\text{\textbf{Attractiveness}}^*$ & Ability to arouse user curiosity and engagement through compelling content. \\
\bottomrule
\end{tabularx}
\end{table}

\section{Implementation Details}
\label{appendix:implementation_details}

To balance exploratory item discovery (cognitive channel) with conversion-driven optimization (utility channel) under limited exposure budgets, we formulate traffic allocation as a constrained quadratic programming problem. This optimization framework dynamically adjusts the exposure proportion of cognitive retrieval items, maximizing overall platform value while ensuring sufficient visibility for novel recommendations.

\paragraph{Problem Formulation}
Let $\mathcal{I} = \{1, 2, \ldots, n\}$ denote the set of candidate items retrieved by the cognitive channel. For each item $i \in \mathcal{I}$, we introduce a decision variable $x_i \in [0, 1]$ indicating the exposure probability of item $i$. The optimization objective is formulated as:
\begin{equation}
\label{eq:qp_objective}
\begin{aligned}
\max_{\mathbf{x}} \quad & \sum_{i \in \mathcal{I}} x_i s_i - \frac{\lambda}{2} \|\mathbf{x}\|^2 \\
\text{s.t.} \quad & \sum_{i \in \mathcal{I}} \left[ x_i o_i + (1 - x_i) \bar{o} \right] \geq \mathcal{C}, \\
& \mathcal{Q} \leq \sum_{i \in \mathcal{I}} x_i \leq \mathcal{P}, \\
& 0 \leq x_i \leq 1, \quad \forall i \in \mathcal{I},
\end{aligned}
\end{equation}
where:
\begin{itemize}[topsep=2pt, itemsep=2pt]
    \item $s_i \in \mathbb{R}^+$ represents the predicted click revenue for item $i$, measuring short-term engagement value;
    \item $o_i \in \mathbb{R}^+$ denotes the predicted conversion revenue (e.g., GMV, transaction value) for item $i$, capturing long-term commercial utility;
    \item $\bar{o} = \frac{1}{|\mathcal{I}|} \sum_{i \in \mathcal{I}} o_i$ is the average conversion revenue across all candidate items;
    \item $\lambda > 0$ is a regularization parameter ensuring strong convexity of the objective function, preventing overfitting to high-revenue items and promoting exposure diversity;
    \item $\mathcal{C} \in \mathbb{R}^+$ is a lower bound on total conversion revenue, ensuring that exploratory recommendations maintain platform-level commercial viability;
    \item $\mathcal{Q}, \mathcal{P} \in \mathbb{R}^+$ ($\mathcal{Q} \leq \mathcal{P}$) define the minimum and maximum number of items eligible for enhanced exposure, controlling the intensity of cognitive exploration.
\end{itemize}

The objective~\eqref{eq:qp_objective} maximizes short-term click revenue while penalizing overly concentrated item selection through the regularization term $\frac{\lambda}{2} \|\mathbf{x}\|^2$. The first constraint ensures that aggregate conversion revenue meets platform commercial targets. The second constraint bounds the number of exposed items within a feasible range, enabling flexible regulation of exploration intensity. The constraint $x_i \in [0,1]$ allows continuous relaxation of binary decisions, facilitating efficient optimization.

\paragraph{Lagrangian Formulation and Analytical Solution}
To solve the constrained optimization problem \eqref{eq:qp_objective}, we construct the Lagrangian function incorporating all constraints via Lagrange multipliers:
\begin{equation*}
\label{eq:lagrangian}
\begin{aligned}
\mathcal{L}(\mathbf{x}, \alpha, \boldsymbol{\beta}, \boldsymbol{\gamma}, \mu, \nu) 
&= \frac{\lambda}{2} \|\mathbf{x}\|^2 - \sum_{i \in \mathcal{I}} x_i s_i \\
&\quad + \alpha \left( \mathcal{C} - \sum_{i \in \mathcal{I}} \left[ x_i o_i + (1 - x_i) \bar{o} \right] \right) \\
&\quad - \sum_{i \in \mathcal{I}} \beta_i x_i + \sum_{i \in \mathcal{I}} \gamma_i (x_i - 1) \\
&\quad + \mu \left( \mathcal{Q} - \sum_{i \in \mathcal{I}} x_i \right) + \nu \left( \sum_{i \in \mathcal{I}} x_i - \mathcal{P} \right),
\end{aligned}
\end{equation*}
where $\alpha, \mu, \nu \geq 0$ are inequality constraint multipliers, and $\beta_i, \gamma_i \geq 0$ enforce box constraints $0 \leq x_i \leq 1$.

Taking the derivative of $\mathcal{L}$ with respect to $x_i$ and applying the KKT optimality conditions:
\begin{equation*}
\frac{\partial \mathcal{L}}{\partial x_i} = \lambda x_i - s_i - \alpha (o_i - \bar{o}) - \beta_i + \gamma_i - \mu + \nu = 0.
\end{equation*}
Rearranging and invoking complementary slackness conditions ($\beta_i x_i = 0$, $\gamma_i (1 - x_i) = 0$), we define $r \triangleq \mu - \nu$ (the net boundary pressure) and obtain the closed-form solution:
\begin{equation*}
\label{eq:solution}
x_i^* = \begin{cases}
1, & \text{if } \; s_i + \alpha (o_i - \bar{o}) + r > \lambda, \\
\frac{s_i + \alpha (o_i - \bar{o}) + r}{\lambda}, & \text{if } \; 0 \leq s_i + \alpha (o_i - \bar{o}) + r \leq \lambda, \\
0, & \text{if } \; s_i + \alpha (o_i - \bar{o}) + r < 0.
\end{cases}
\end{equation*}

\textbf{Interpretation.} The exposure decision for item $i$ is determined by the composite score $h_i \triangleq s_i + \alpha (o_i - \bar{o}) + r$, which aggregates three components:
\begin{itemize}[topsep=2pt, itemsep=2pt]
    \item \textbf{Click Revenue} $s_i$: Captures immediate user engagement value.
    \item \textbf{Conversion Premium} $\alpha (o_i - \bar{o})$: Weights items with above-average conversion revenue, where $\alpha \geq 0$ modulates the trade-off between click-oriented and conversion-oriented optimization.
    \item \textbf{Budget Pressure} $r = \mu - \nu$: Dynamically adjusts exposure intensity based on constraint tightness. When total exposure approaches $\mathcal{Q}$ (lower bound), $\mu$ increases, raising $r$ to admit more items; conversely, when approaching $\mathcal{P}$ (upper bound), $\nu$ increases, lowering $r$ to restrict admission.
\end{itemize}

Items satisfying $h_i > \lambda$ receive full exposure ($x_i = 1$), those with $0 \leq h_i \leq \lambda$ receive fractional exposure proportional to $h_i / \lambda$, and items with $h_i < 0$ are excluded.

\paragraph{Online Deployment and Practical Simplification}
Due to stringent latency constraints in production systems, we adopt a binarized exposure strategy that simplifies the fractional exposure case. Specifically, items with $0 < h_i \leq \lambda$ (intermediate scores) are treated as $x_i = 0$ and excluded from exposure, effectively implementing a hard threshold policy:
\begin{equation*}
\label{eq:simplified_solution}
x_i^* = \begin{cases}
1, & \text{if } \; h_i > \lambda, \\
0, & \text{otherwise}.
\end{cases}
\end{equation*}
This simplification eliminates the need for strict solution (\textit{e.g.}, branch and bound algorithms) at inference time, reducing computational overhead.
This quadratic programming framework provides a principled mechanism for harmonizing cognitive exploration with utility-driven exploitation, achieving sustainable recommendation ecosystem growth through flexible traffic allocation.

\end{CJK*}
\end{document}